\documentclass[12pt]{article}

\begin{document}


\newcommand{\bb}{\begin{equation}}
\newcommand{\ee}{\end{equation}}
\newcommand{\bbb}{\begin{eqnarray}}
\newcommand{\eee}{\end{eqnarray}}
\newcommand{\vc}[1]{\mbox{$\vec{{\bf #1}}$}}
\newcommand{\mc}[1]{\mathcal{#1}}
\newcommand{\del}{\partial}
\newcommand{\lk}{\left}
\newcommand{\ave}[1]{\mbox{$\langle{#1}\rangle$}}
\newcommand{\re}{\right}
\newcommand{\pd}[1]{\frac{\del}{\del #1}}
\newcommand{\pdd}[2]{\frac{\del^2}{\del #1 \del #2}}
\newcommand{\Dd}[1]{\frac{d}{d #1}}
\newcommand{\sech}{\mbox{sech}}
\newcommand{\pref}[1]{(\ref{#1})}

\input{epsf}


\def\journal#1&#2(#3){\unskip, \sl #1\ \bf #2 \rm(19#3) }
\def\andjournal#1&#2(#3){\sl #1~\bf #2 \rm (19#3) }
\def\nextline{\hfil\break}

\def\ie{{\it i.e.}}
\def\eg{{\it e.g.}}
\def\cf{{\it c.f.}}
\def\etal{{\it et.al.}}
\def\etc{{\it etc.}}

\def\sst{\scriptscriptstyle}
\def\tst#1{{\textstyle #1}}
\def\frac#1#2{{#1\over#2}}
\def\coeff#1#2{{\textstyle{#1\over #2}}}
\def\half{\frac12}
\def\hf{{\textstyle\half}}
\def\nth{$n^{\rm th}$}
\def\ket#1{|#1\rangle}
\def\bra#1{\langle#1|}
\def\vev#1{\langle#1\rangle}
\def\d{\partial}

\def\inbar{\,\vrule height1.5ex width.4pt depth0pt}
\def\IC{\relax\hbox{$\inbar\kern-.3em{\rm C}$}}
\def\IR{\relax{\rm I\kern-.18em R}}
\def\IP{\relax{\rm I\kern-.18em P}}
\def\Z{{\bf Z}}
\def\R{{\bf R}}
\def\One{{1\hskip -3pt {\rm l}}}
\def\np#1#2#3{Nucl. Phys. {\bf B#1} (#2) #3}
\def\npb#1#2#3{Nucl. Phys. {\bf B#1} (#2) #3}
\def\pl#1#2#3{Phys. Lett. {\bf #1B} (#2) #3}
\def\plb#1#2#3{Phys. Lett. {\bf #1B} (#2) #3}
\def\prl#1#2#3{Phys. Rev. Lett. {\bf #1} (#2) #3}
\def\physrev#1#2#3{Phys. Rev. {\bf D#1} (#2) #3}
\def\prd#1#2#3{Phys. Rev. {\bf D#1} (#2) #3}
\def\annphys#1#2#3{Ann. Phys. {\bf #1} (#2) #3}
\def\prep#1#2#3{Phys. Rep. {\bf #1} (#2) #3}
\def\rmp#1#2#3{Rev. Mod. Phys. {\bf #1} (#2) #3}
\def\cmp#1#2#3{Comm. Math. Phys. {\bf #1} (#2) #3}
\def\cqg#1#2#3{Class. Quant. Grav. {\bf #1} (#2) #3}
\def\mpl#1#2#3{Mod. Phys. Lett. {\bf #1} (#2) #3}
\def\ijmp#1#2#3{Int. J. Mod. Phys. {\bf A#1} (#2) #3}
\def\jmp#1#2#3{J. Math. Phys. {\bf #1} (#2) #3}
\catcode`\@=11
\def\slash#1{\mathord{\mathpalette\c@ncel{#1}}}
\overfullrule=0pt
\def\AA{{\cal A}}
\def\BB{{\cal B}}
\def\CC{{\cal C}}
\def\DD{{\cal D}}
\def\EE{{\cal E}}
\def\FF{{\cal F}}
\def\GG{{\cal G}}
\def\HH{{\cal H}}
\def\II{{\cal I}}
\def\JJ{{\cal J}}
\def\KK{{\cal K}}
\def\LL{{\cal L}}
\def\MM{{\cal M}}
\def\NN{{\cal N}}
\def\OO{{\cal O}}
\def\PP{{\cal P}}
\def\QQ{{\cal Q}}
\def\RR{{\cal R}}
\def\SS{{\cal S}}
\def\TT{{\cal T}}
\def\UU{{\cal U}}
\def\VV{{\cal V}}
\def\WW{{\cal W}}
\def\XX{{\cal X}}
\def\YY{{\cal Y}}
\def\ZZ{{\cal Z}}
\def\lam{\lambda}
\def\eps{\epsilon}
\def\vareps{\varepsilon}
\def\underrel#1\over#2{\mathrel{\mathop{\kern\z@#1}\limits_{#2}}}
\def\lapprox{{\underrel{\scriptstyle<}\over\sim}}
\def\lessapprox{{\buildrel{<}\over{\scriptstyle\sim}}}
\catcode`\@=12
\def\sdtimes{\mathbin{\hbox{\hskip2pt\vrule height 4.1pt depth -.3pt width
.25pt \hskip-2pt$\times$}}}
\def\bra#1{\left\langle #1\right|}
\def\ket#1{\left| #1\right\rangle}
\def\vev#1{\left\langle #1 \right\rangle}
\def\det{{\rm det}}
\def\tr{{\rm tr}}
\def\mod{{\rm mod}}
\def\sinh{{\rm sinh}}
\def\cosh{{\rm cosh}}
\def\sgn{{\rm sgn}}
\def\det{{\rm det}}
\def\exp{{\rm exp}}
\def\sh{{\rm sh}}
\def\ch{{\rm ch}}

\def\bh{{\sst BH}}
\def\lc{{\sst LC}}
\def\pr{{\sst \rm pr}}
\def\cl{{\sst \rm cl}}
\def\D{{\sst D}}
\def\st{\scriptstyle}
\def\aleff{{\alpha'_{\rm eff}}}
\def\lpl{\ell_{{\rm pl}}}
\def\str{{s}}
\def\lstr{\ell_{{\str}}}
\def\gstr{g_\str}
\def\mm{{\rm \sst M}}
\def\osc{{\rm\sst osc}}
\def\gym{g_{Y}}
\def\aa{{\sst (a)}}
\def\lpla{{\bar \ell}_{{\rm pl}}}
\def\lstra{{\bar\ell}_{{\str}}}
\def\Ma{{\overline {\rm M}}}

\begin{titlepage}
\rightline{EFI-98-47}

\rightline{hep-th/9809061}

\vskip 3cm
\centerline{\Large{\bf Black Holes and the SYM Phase Diagram}}

\vskip 2cm
\centerline{Miao Li\footnote{\texttt{mli@theory.uchicago.edu}},~~~ 
Emil Martinec\footnote{\texttt{ejm@theory.uchicago.edu}}, ~~and~~~ 
Vatche Sahakian\footnote{\texttt{isaak@theory.uchicago.edu}}}
\vskip 12pt
\centerline{\sl Enrico Fermi Inst. and Dept. of Physics}
\centerline{\sl University of Chicago}
\centerline{\sl 5640 S. Ellis Ave., Chicago, IL 60637, USA}

\vskip 2cm

\begin{abstract}

Making combined use of the Matrix and Maldacena conjectures,
the relation between various thermodynamic transitions
in super Yang-Mills (SYM) and supergravity is clarified. The thermodynamic
phase diagram of an object in DLCQ M-theory 
in four and five non-compact space dimensions is constructed;
matrix strings, matrix black holes,
and black $p$-branes are among the various phases. 
Critical manifolds are characterized by the principles 
of correspondence and longitudinal localization, 
and a triple point is identified. 
The microscopic dynamics of the Matrix string near two of
the transitions is studied; we identify a signature of black hole
formation from SYM physics.

\end{abstract}

\end{titlepage}
\newpage
\setcounter{page}{1}
\section{Introduction and summary}

The thermodynamic phase structure of a theory is an excellent
probe into the underlying physics. Transitions among different phases 
reflect the dynamics via the stability or metastability of
various configurations, while order parameters often 
characterize global properties.
M/String theory is no exception in this regard.

In particular, two thermodynamic transition mechanisms
in M/String theories have recently been a focus of the literature.
One example occurs when the curvature near the horizon of
a supergravity solution becomes of the order of the string scale; 
the state becomes `stringy', and
acquires an alternative string theoretical description,
either by a perturbative string or by 
supersymmetric Yang-Mills (SYM) D-brane dynamics~\cite{CORR1}.
This metamorphosis might be regarded as
a phase transition in the embedding theory, 
and is known as the correspondence principle.
A second transition mechanism is associated with the mechanics
of localizing a state in a compact direction. 
Of particular interest in discrete light-cone quantization (DLCQ)
is the localization effect in the longitudinal direction
$R_+$ (or $R_{11}$ in the 
infinite momentum frame (IMF))~\cite{BFKS1,BFKS2,BFKS3,HORMART}. 
Particularly, a state with fixed rest mass $M$
and $N$ units of DLCQ momentum satisfies the condition 
\bb\label{rwrtm}
R_+<\frac{N}{M}\equiv q^{-1}\ .
\ee
If the system characterizes an object of size $r_0$, 
then we need $r_0<R_+$ to localize the object. 
For example, for a black hole satisfying
the equation of state $M r_0=S$, we need
\bb\label{ns}
N>S\ ,
\ee
otherwise the black hole fills the longitudinal direction
and becomes a black string.
\footnote{This intuitive argument ignores the effects of 
gravity.  The momentum of an object
back-reacts on the nearby geometry and
in particular changes the available proper longitudinal volume;
such effects do not 
affect the conclusion \pref{ns}, however \cite{HORMART}.}
We will call the transition
at $N\sim S$ a localization transition.
Other geometrical, but non-longitudinal, 
effects of this sort may also be expected~\cite{SUSSGW}.

If there existed a single framework -- one description that 
realizes the different phases of M-theory -- 
then this theory should exhibit
the critical phenomena associated with these various transitions.
The Matrix theory conjecture proposes such a framework: 
$U(N)$ SYM on a torus is supposed to manifest
a rich structure of M/String theory phases 
such as black holes, strings and D-branes~\cite{MAT1}.
SYM thermodynamics is then endowed with a cornucopia of
critical behaviors. Field theoretically,
transitions between SYM phases are possible as functions
of the size and shape of the torus, 
the YM coupling, the temperature, and the rank of the
gauge group. 

A complementary recent conjecture of Maldacena~\cite{MALDA1,MALDA2}
provides us with the tools
to study SYM thermodynamics in regimes 
previously considered intractable.
It states that the macroscopic physics of M/string theory 
in the vicinity of some large charge source
(a regime accurately described by supergravity), 
is equivalent to that of super Yang-Mills.
Finite temperature SYM physics acquires in certain regimes 
a geometrical description, that of the near horizon region
of near-extremal supergravity solutions. 
RG flow is mapped onto transport in the geometry about the horizon;
correlation functions in the SYM probe different
distances from the horizon as one changes the separation of operator
insertions relative to the correlation length 
(thermal wavelength) in the SYM.

Our plan is to use the Maldacena conjecture, 
along with the interpretation of the SYM
physics from the Matrix theory perspective, 
to piece together the phase diagram
of an object in DLCQ M-theory.
In parallel, we will end up making statements
about the critical behavior of SYM thermodynamics on the torus
well into non-perturbative field theory regimes.

We focus on type IIA theory on $T^p$ with $p=4,5$ 
(4+1d and 5+1d SYM theory).
One reason for this is that
gravitational interactions are
longer-range, and sufficiently strong
to be important for state transitions,
in lower dimensions (higher $p$).
Also, the cases with $p=1$ and $p=2$ require slightly more effort;
and the case $p=3$ is conformal -- 
the SYM coupling $g_Y$ is dimensionless, so that
quantitative control is required for some questions to be addressed. 
We confine this report to a more qualitative sketch
of the phase diagram; for example, we ignore 
numerical coefficients in state equations.
We hope to return to a discussion of the
situation for $p\le 3$ elsewhere.  In particular,
the cases $p=3,2$ are relevant to both matrix theory
and string theory in anti-de Sitter space, and
thus their phase diagrams should be rather interesting.

\begin{figure}[p]
\epsfxsize=13cm \centerline{\leavevmode \epsfbox{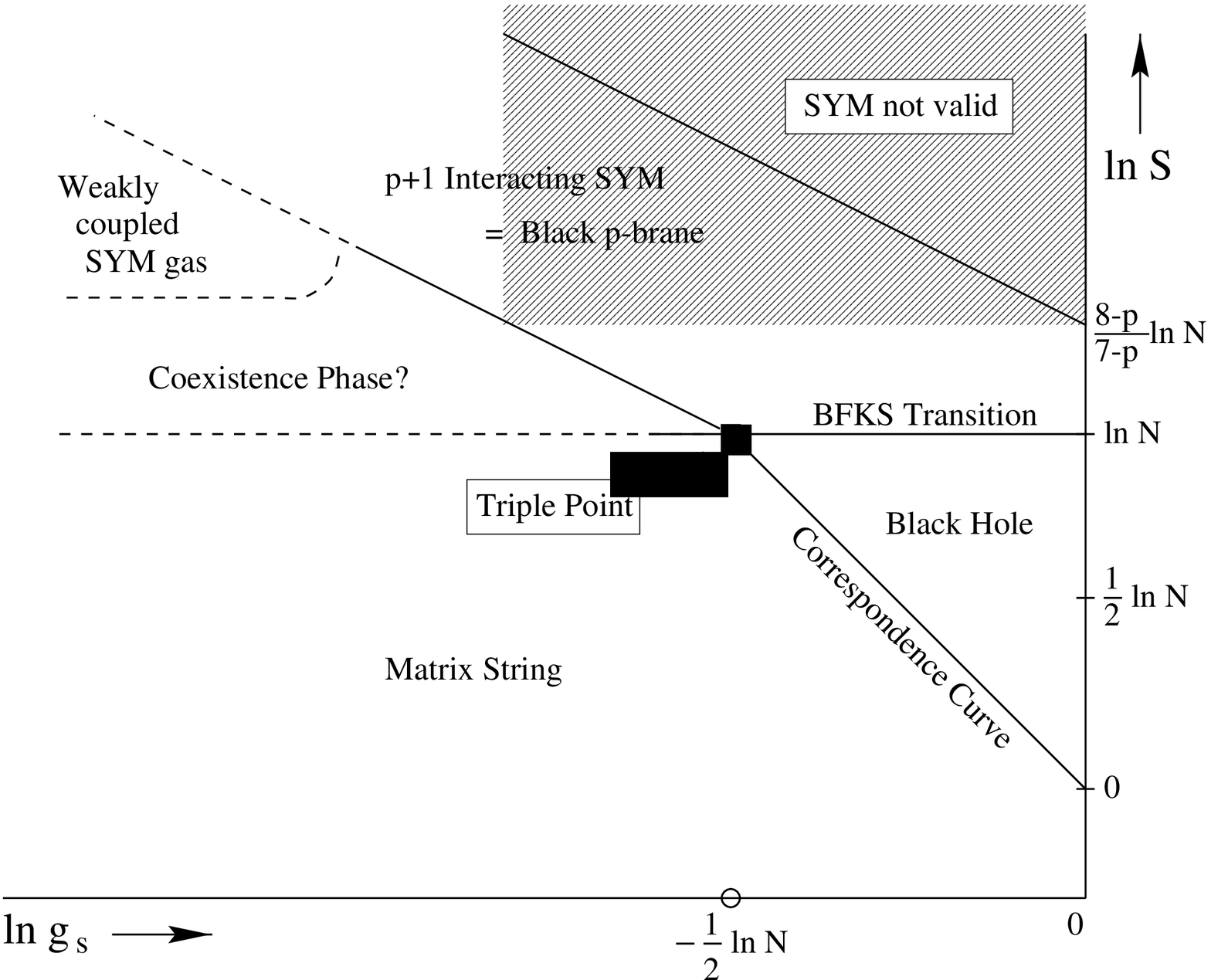}}
\caption{\sl The proposed thermodynamic phase diagram for the $p+1$d SYM
on the torus, or the DLCQ IIA theory, obtained by tracking an object in
Matrix theory.  On the horizontal axis is the IIA string coupling,
which is the aspect ratio of the SYM torus.  The vertical
axis is the density of states of the object.
}
\label{fig3}
\end{figure}

Our main conclusions come in two pieces. The first consists of
an overview of previous
observations~\cite{CORR1,BFKS2,HORMART,MALDA2,SUSSGW}, 
put in a new unifying perspective.
Figure~\ref{fig3} summarizes the situation. 
It is the phase diagram traced out by a {\it single object} in
matrix theory on $T^4$ or $T^5$.
We are assuming that the different states we track
are characterized by long enough lifetimes so that it makes
sense to describe them thermodynamically, as (meta)stable phases.
In super Yang-Mills,
one has in mind starting the system with all scalar field
vacuum expectation values
bounded in some appropriately small region, such that the interactions
sustain a long-lived cohesive state.

In the figure,
the limit of validity of the SYM description for the DLCQ string theory 
is determined by the upper right curve.  
In the shaded region, the theory is sufficiently
strongly coupled at the scale of the temperature that it
is not accurately described by super Yang-Mills theory;
rather, one must pass to the six-dimensional (2,0) theory \cite{BRS}
for $p=4$, or the ill-understood `little string' theory 
\cite{SEIBLITTLE,DVV5D}
for $p=5$.  We will see that the dynamics of interest
to us occurs outside this region.
We identify several phases in SYM on the torus; 
a black hole phase, a string phase,
a phase of $p+1$-dimensional strongly interacting SYM, 
and perhaps a `coexistence phase'
of a Matrix string with SYM vapor.  
On the upper left part, the system evaporates into 
a weakly coupled SYM gas, over sufficiently short time scales
that one cannot think of the ensemble as that of a single object
in spacetime.
There is a `triple point', a thermodynamic
critical point of the DLCQ string theory where the three 
transition manifolds coincide.

A brief description of the physics of the diagram is as follows: 
In type II DLCQ string theory on $T^p$, with $p=4,5$, 
there exists a (longitudinally wrapped) D$p$-brane phase; 
it is unstable at the BFKS point (the horizontal line
at $N\sim S$ in the diagram) to the formation
of a black hole because of longitudinal localization effects.
Along another critical curve (the diagonal line above $N\sim S$), 
the D$p$ brane freezes its strongly
coupled excitations onto a single direction of the torus, 
making a transition to a perturbative string through the 
correspondence principle. 
In this regime, the thermodynamics is that of a 
near-extremal fundamental (IIB) string supergravity solution,
with curvature at the horizon becoming of order the string scale. 
The correspondence mechanism also applies on the other side
of the BFKS transition; in this case, a matrix black hole
makes a transition to a matrix string when it acquires
string scale curvature at the horizon.
A coexistence phase, where both matrix string and SYM gas
excitations contribute strongly to the thermodynamics,
may exist in the region indicated on the diagram; 
this depends on the extent to which the object persists
long enough to treat it using the methods of 
equilibrium thermodynamics.

\begin{figure}[t]
\epsfxsize=7cm \centerline{\leavevmode \epsfbox{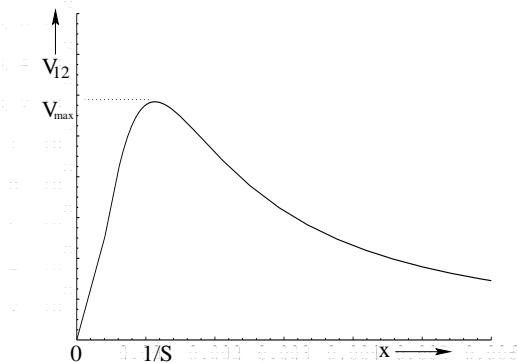}}
\caption{\sl The string self-interaction potential as a function of 
relative separation $x$ along the string, for $p=4,5$.}
\label{potl}
\end{figure}

Our second set of results concerns the dynamics that
leads to the correspondence transition,
and is summarized by Figure~\ref{potl}.
The plot depicts the mutual gravitational interaction energy
between a pair of points on a typical (thermally excited)
macroscopic Matrix string, as a function of 
the world-sheet distance $x$ along the string separating the two points.
This potential governs the
dynamics of the Matrix string near
a black hole or black brane transition,
as it is approached from the weak coupling side. 
A bump in the potential occurs at the thermal wavelength $N/S$ 
for $p=4,5$ (five or four noncompact spatial directions);
in these dimensions, the correspondence transition
to a black hole is indeed caused by the
string's self-interaction, as discussed in~\cite{CORR2}. 
For smaller $p$ (more noncompact spatial directions),
there is no bump; similarly, in \cite{CORR2}
the self-interactions could not cause a spontaneous
collapse to a black object.
We will see that the height of the bump is proportional to the
gravitational coupling, such that it `confines'
excitations of the string on the strong-coupling
side of the correspondence transition.

This result supports a suggestion \cite{LIMART,MIAO} to describe
the black hole phase as clustered Matrix SYM 
excitations of size $N/S$.  These correlated
clusters were invoked in order that the
object with $N>S$ can be localized in the longitudinal direction. 
Such a localization necessarily 
involves the longitudinal momentum physics 
of Matrix theory.  We find that a plausible argument for the dynamics 
with this potential gives the previously identified correspondence curves
as the boundaries of validity of the matrix string phase.
For $N<S$, one finds the transition to the interacting $p+1$
SYM phase shown in Figure~\pref{fig3};
while for $N>S$, one finds the transition to the matrix
black hole phase.  Accounting for the latter transition requires taking into 
consideration longitudinal momentum transfer effects as in~\cite{LIMART};
we justify this by a string theory amplitude calculation 
involving winding number exchange in a dual picture.
We thus conclude that we have identified the characteristics of the
microscopic mechanism of black hole formation 
from the SYM point of view.

The plan of the presentation is as follows:
in section 2, we review the two conjectures 
(Matrix and Maldacena) we will use in the analysis of the
phase diagram.
In section 3, we bring together 
previous observations with some new ones to
map out the phase diagram for the DLCQ Matrix string. 
Section 4 extends our arguments at the triple point
to the cases of singly and doubly charged black holes.
We also discuss a toy
mechanism for clustering of SYM excitations 
for the singly charged case, at the BFKS point. 
Section 5 discusses the self-interaction of the Matrix string, 
the identification of the bump potential and comments about its dynamics.
We outline in the Appendices the calculation of the potential, and a
scattering amplitude calculation relevant the issue of longitudinal 
momentum transfer physics.

As we were finalizing the manuscript, 
a paper discussing related issues 
\cite{RABIN}
came to our attention.

\section{A couple of conjectures}

\subsection{The Matrix conjecture}

A convenient way to summarize the Matrix theory conjecture
is to say that DLCQ M-theory on $T^p$ with $N$ units of
longitudinal momentum is a particular regime
of an auxiliary `$\Ma$-theory' which freezes
the dynamics onto a subsector of that theory.
Consider such an $\Ma$-theory, with eleven-dimensional
Planck scale $\lpla$ (which we denote ($\Ma$,$\lpla$))
on a $p+1$d dimensional torus of radii ${\bar{R}}_i$, $i=1\ldots p$, and
$\bar{R}$ the `M-theory circle' of reduction to type $\overline{\rm IIA}$
string theory, in the limiting regime
\bb \label{limitlp}
\lpla\rightarrow 0,\ \ \ 
\mbox{with } x\equiv \frac{\lpla^2}{\bar{R}}\ \ \ \mbox{and }
y_i\equiv \frac{\lpla}{\bar{R}_i}\ \ \ \mbox{fixed} ,
\ee
and $N$ units of KK momentum along $\bar{R}$. 
It is proposed that~\cite{MAT1,MAT2}
\begin{itemize}
\item 
This theory is equivalent to an ($M$,$\lpl$) 
theory on the DLCQ background
we denote by $D^{1,1}\times T^p \times \R^{9-p}$, 
where $D^{1,1}$ is a $1+1$ dimensional subspace compactified
on a lightlike circle of radius $R_+$, and the torus $T^p$ has
radii $R_i$ ($i=1\ldots p$).
The map between the two theories is given by
\bb \label{map}
x=\frac{\lpl^2}{R_+},\ \ \ 
y_i=\frac{\lpl}{R_i} ,
\ee
with $N$ units of momentum along $R_+$.
\item
The dynamics of
the $\Ma$ theory in the above limit can be described
by a subset of its degrees of freedom, that of
$N$ D0 branes of the $\overline{\rm IIA}$ theory, 
up to a certain UV cutoff.

\end{itemize}

The two propositions above, in conjunction, 
are referred to as the Matrix conjecture~\cite{MAT1,MAT2}.

T-dualizing on the $\bar{R}_i$'s, we describe the D0 brane physics 
by the $p+1$d SYM of N D$p$ branes
wrapped on the dualized torus. We remind the reader
of the dictionary needed in this process
\bbb 
\bar{R}=&~\bar{g}_s' \lstra\ \ \ 
\lpla^3=\bar{g}_s' \lstra^3\ ,\nonumber\\
\bar{g}_s=&~\bar{g}_s' \frac{\lstra^p}{\Pi \bar{R}_i}\ \ \ 
\Sigma_i=\frac{\lstra^2}{\bar{R}_i}\ .\label{dictionary1}
\eee
The first line is the $\Ma-\overline{\rm IIA}$ 
relation, the second that of T-duality.
The limit~\pref{limitlp} then translates in the new variables to
\bb \label{limitym}
  \bar{\alpha}'\rightarrow 0\quad, \quad \quad 
  \mbox{with }\ \  g_Y^2=(2 \pi)^{p-2} 
	\bar{g}_s {{\bar\alpha}}^{\prime(p-3)/2}\ \ \ 
  \mbox{and } \Sigma_i\ \ \ \mbox{fixed}\ ,
\ee
where the nomenclature 
$g_Y^2$ and $\Sigma_i$ refers to the coupling and
radii of the corresponding $p+1$d
$U(N)$ SYM theory, whenever it is well defined in this limit,
\ie\ for $p\le 3$. 

For $p>3$, we see from~\pref{limitym} that $\bar{g}_s\rightarrow \infty$, 
the dilaton at infinity diverges
(\ie\ in the UV of the field theory, according to Maldacena's
conjecture~\cite{MALDA1}); this is a statement of the
non-renormalizability of the corresponding SYM: New physics
sets in the UV. For $p=4$,
the D-branes physics 
is the IR limit of the six-dimensional (2,0) theory;
while for $p=5$, 
it is that of a weakly coupled 
IIB NS5 brane~\cite{WHYSEIB,ASHOKE}. 
We ignore hereafter all cases with $p>5$. 
In summary, only at low enough energies the 
4+1d and 5+1d SYM yield a proper coarse-grained description 
of the needed dynamics.

\subsection{The Maldacena conjecture}

It is proposed~\cite{MALDA1}
that in the limit~\pref{limitym}, one can identify the
physics of the SYM QFT at different energy scales with the supergravity
solution that is cast by the branes, 
whenever such a solution is well defined.
One is to identify string theory excitations 
of the supergravity background with those of the QFT; 
this is essentially a correspondence 
between closed and open string dynamics.

Here, we will study finite temperature physics.
We will therefore make use of the 
thermodynamic version of the above statement,
which identifies the finite temperature vacuum of the SYM
describing $N$ D$p$ branes with the geometry of the 
horizon region of the near extremal supergravity solution 
the branes cast about them, 
whenever such a solution is sensible~\cite{MALDA2}. 
In particular, we can extract
the thermodynamics of SYM at finite
temperature in its non-perturbative regimes.
Correlation functions in the SYM probe different
distances from the horizon in the supergravity solution
as one changes the separation of operator
insertions relative to the SYM correlation length 
(thermal wavelength);
coarse graining the SYM theory to lower energies corresponds to moving
towards to the center of the supergravity solution, up 
until the near extremal horizon~\cite{SUSSHOLO}. 

Our strategy will be to use
the Maldacena conjecture as a tool, to study the thermodynamics
of Matrix strings and black holes; and conversely,
to learn about the phase diagram of supersymmetric
Yang-Mills theory on the torus.

\section{A thermodynamic road-map}

\subsection{Preliminaries}

A DLCQ IIA theory descends from the DLCQ $M$
theory described above; we choose string scale compactification
\bb \label{IIAsetup}
R_i \sim \lstr\ \ \
\mbox{for } i=1\ldots p-1\ ,
\ee
with
\bb \label{IIAsetup2}
R_p=g_s \lstr\ \ \ \lpl^3=g_s \lstr^3\ ,
\ee
and a perturbative IIA regime
\bb \label{IIAsetup3}
g_s< 1\ .
\ee
We can in principle relax~\pref{IIAsetup} at the expense of introducing new
state variables, and a more complicated (and richer) phase diagram; 
for simplicity, we will stick to this `IIA regime'. Using the equations
in the previous section, we can write the dictionary between our IIA
theory and the Matrix SYM
\bbb 
g_Y^2&=&~(2 \pi)^{p-2} (a g_s)^{p-3}\ ,\nonumber\\
\Sigma_i&=&~g_s a\ \ \ \mbox{for } i=1\ldots p-1\ ,\nonumber\\
\Sigma_p&=&~a\ ,\label{maindict}\\
V&\equiv &~\Sigma_i^{p-1} \Sigma_p=g_s^{p-1} a^p\ ,\nonumber
\eee
with
\bb
a\equiv \frac{\alpha'}{R_+}\ .
\ee
We chose~\pref{IIAsetup3} so that we have
$\Sigma_i<\Sigma_p$, simplifying our analysis later.

We study finite temperature physics of
this IIA theory with the finite temperature vacuum of the corresponding SYM.
As mentioned in the introduction, we confine our analysis
to $p=4$ and $p=5$.

\subsection{Validity of SYM}

Given that we are working with Matrix theory on $T^4$ and $T^5$, the first
question that must be addressed concerns the validity of the
description, given that SYM 4+1d and 5+1d are non-renormalizable.
New degrees of freedom are required to make sense of the SYM
dynamics as we probe it in the UV,
\ie\ as we navigate outward 
in the corresponding supergravity solution.
These new degrees of freedom for SYM 4+1d and 5+1d are associated with 
the onset of strong coupling dynamics; 
the validity of the theories at different energy scales
is then determined by looking at the 
size of the dilaton vev at different locations
in the supergravity solution.
For finite temperatures, 
physics at the thermal wavelength of the SYM is identified with 
physics at the horizon of the near extremal solution~\cite{MALDA2,KLEBTSEYT}
\bb \label{near}
ds^2_{\rm hor}=\alpha' \lk( \frac{U^{(7-p)/2}}{g_Y\sqrt{d_p N}}
d\sigma_i^2
+g_Y\sqrt{d_pN} U^{(p-3)/2} d\Omega_{8-p}^2 \re )\ .
\ee
We are looking at a fixed time and radial slice; 
the radial variable is $U$,
the $\sigma_i$'s are coordinates
along the brane with identification $\sigma_i\sim
\sigma_i+\Sigma_i$; $d_p$ is a numerical coefficient; and $U_o$ is the
location of the horizon, related to the SYM entropy~\cite{MALDA2}
\bb \label{U0}
U_o^{p-9}\sim (g_Y^2)^{-3} S^{-2} N V^2\ .
\ee
The dilaton vev is
\bb \label{dilaton}
e^\phi=(2\pi)^{2-p} g_Y^2 \lk( \frac{g_Y^2 d_p N}{U^{7-p}} \re)^{(3-p)/4}\ .
\ee
The finite temperature vacuum of the 4+1d and 5+1d
SYM is a valid thermodynamic 
description of the DLCQ IIA theory (by the two conjectures 
stated earlier) when
\bb \label{dillim}
\lk.e^\phi \re|_{U_o}\ll 1 \Rightarrow S \ll N^{\frac{8-p}{7-p}} g_s^{-1}\ .
\ee
Note that this is a purely geometric statement, in terms of 
the horizon area and string coupling; 
it will be seen to be insensitive 
to finite size effects due to the
transverse torus. We then choose to work on
a two dimensional cross section in the $S$-$g_s$ plane
of the thermodynamic phase diagram, with fixed $N\gg 1$. 
In principle, one is to take the thermodynamic limit $N\rightarrow \infty$
with $N/S$ fixed, to see criticality; transition between phases 
at finite $N$ discussed here are smooth crossovers. 
It is expected that, in the infinite $N$ limit, 
the physics tends to the appropriate critical behavior.

\begin{figure}[t]
\epsfxsize=7cm \centerline{\leavevmode \epsfbox{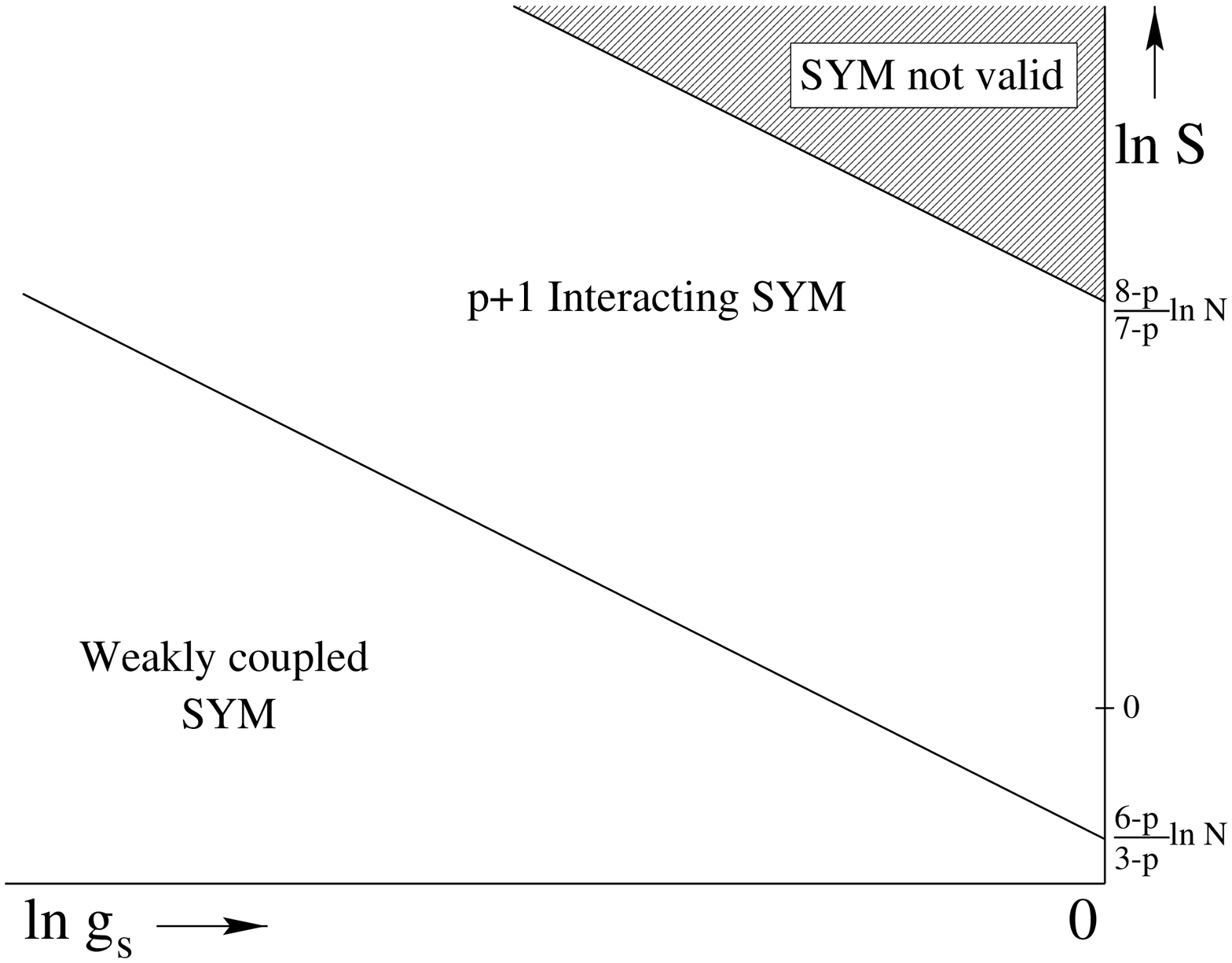}}
\caption{\sl  The entropy $S$ versus $g_s$ phase diagram showing the
region of validity of the SYM description, and the boundary between
the free and interacting phases, ignoring finite size effects. We assume
$N,S \gg 1$, and $g_s< 1$, and fix $N$ for a given diagram.}
\label{fig1}
\end{figure}

Let us for a while ignore the effects of the transverse torus.
In the regime where the curvature of the 
supergravity solution at the horizon 
is less than the string scale, 
\bb \label{curv}
S \gg N^{\frac{p-6}{p-3}} g_s^{-1}\ ,
\ee
the SYM statistical mechanics
obeys the equation of state~\cite{MALDA2,KLEBTSEYT,DUFF}
\bb \label{eosint}
E_{int}^{p-9}\sim S^{2(p-7)} (g_Y^2)^{p-3} N^{7-p} V^{5-p}\ .
\ee
Beyond this regime, 
we have weakly coupled SYM, roughly a free gas of $N^2$ gluons.
These statements are graphically summarized in Figure~\ref{fig1}.

Our assumption that
finite size effects, due to the compactification of the $\sigma_i$
variables, are of no relevance is the statement that, for high enough
temperature (\ie\ for thermal wavelengths short enough with respect
to the effective sizes of the torus), the local
thermodynamics is similar to that of the uncompactified case. The rest of
this section is the study of the breakdown of this regime;
we would like to paint over Figure~\ref{fig1} new phases arising due to
the compactification of the background. In particular, we will argue that
the geometry describing the interacting $p+1$d SYM phase gets modified 
due to the two transition mechanisms outlined in the Introduction;
consequently, the correspondence line of equation~\pref{curv} is changed.

\subsection{Finite size effects}

\begin{figure}[t]
\epsfxsize=7cm \centerline{\leavevmode \epsfbox{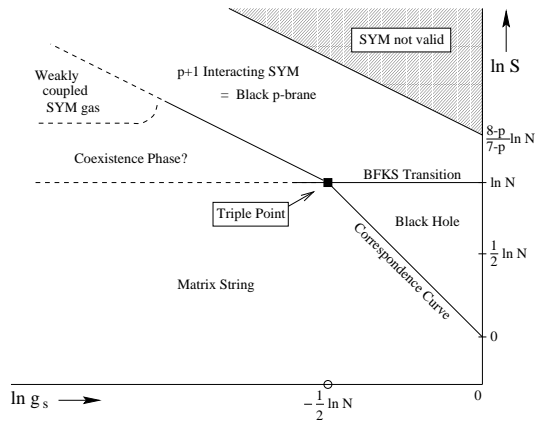}}
\caption{\sl For the convenience of the reader,
we reproduce Figure~\pref{fig3}: 
The proposed thermodynamic phase diagram for the $p+1$d SYM
on the torus, \ie\ the DLCQ IIA theory.}
\label{newfig3}
\end{figure}

A finite size effect in the supergravity regime
that was determined in the work 
of~\cite{BFKS1,BFKS2,BFKS3,HORMART} is the effect of the DLCQ
radius $R_+$ on the geometry. 
We saw from equation~\pref{ns} that a black hole
is localized longitudinally when $N>S$. 
The black hole equation of state is
\bb \label{eosbh}
E^{p-9}_{bh}\sim E_{int}^{p-9} \lk(\frac{N}{S}\re)^2\ .
\ee
The process of minimizing the Gibbs energies between the black hole and
interacting SYM phases yields a black hole phase
as in Figure~\ref{newfig3}, independent of $p$, 
which is the BFKS observation~\cite{BFKS1,BFKS2,BFKS3};
but we now see that the Maldacena conjecture justifies the procedure.

The Schwarzschild black hole geometry will become stringy when
its curvature near the
horizon becomes of the order of the string scale; the emerging state
is a Matrix string in the Matrix conjecture language, \ie\ a $1+1$ state with 
$Z_N$ holonomy on $\Sigma_p$.
Minimizing the Gibbs energy between the Matrix string and Matrix black
hole phases leads to the Horowitz-Polchinski 
correspondence curve~\cite{CORR1,CORR2}
\bb \label{corr}
S\sim g_s^{-2}\ .
\ee
This is a statement independent of $N$ and $p$. 
At $g_c\sim N^{-1/2}\sim N_\osc^{-1/4}$, 
where $N_\osc$ is the string oscillator level,
there exists an interesting critical point.

We next deal with finite size effects in the interacting $p+1$d SYM phase
which are due to the other radii $\Sigma_i$. This is problematic
since, unlike the free case, the strongly coupled SYM may acquire at
finite temperature a nontrivial vacuum; for example, 
a vacuum characterized by a holonomy sewing branes together. 
It is in such a regime that
Susskind observed for the case $p=3$ that the effective size of the
SYM box is bigger than $\Sigma_i$~\cite{SUSSGW}. 
In the spirit of Susskind's transition, and inspired by the existence
of a Matrix string phase on the left 
of the phase diagram, we suggest that
the finite size effect can be probed by comparing the Gibbs energies
of the interacting $1+1$d and $p+1$d phases. Using equation~\pref{eosint},
we get a critical line at
\bb \label{spec1}
S\sim \sqrt{N} g_s^{-1}\ ,
\ee
independent of $p$ and matching\footnote{Figuratively speaking; 
we have dropped numerical coefficients in this analysis; strictly speaking,
this critical `point' may be a manifold of dimension greater than $0$.} 
onto the `triple point'
of correspondence.  Even so, for these values of $S$, $N$ and $g_s$, 
the $1+1$d SYM is not described by the supergravity solution
whose equation of state we used.
The analogue of Figure~\ref{fig1} with $p=1$
has the correspondence curve on the strong coupling side of the
line where $e^\phi=1$; in other words,
the D string supergravity solution is strongly
coupled, as noted in~\cite{MALDA2}.  However, 
the S-dual is the weakly coupled supergravity solution of a
fundamental IIB string source;
its equation of state is the same as the one used above, given that the
entropy is to be calculated in the Einstein frame.
The curvature at the horizon of the S-dual solution
becomes of order the string scale at precisely~\pref{spec1} \cite{MALDA2},
beyond which a Matrix string description emerges. We can further check
the correctness of this conclusion by
matching the $p+1$d interacting SYM gas
equation of state with that of the Matrix string, the latter being the
dominant phase on the other side of this correspondence curve.
The result is again~\pref{spec1}.
We conclude that the $p+1$d interacting SYM makes
a transition to a Matrix String at~\pref{spec1}. 
This is shown in Figure~\ref{newfig3}.

From the supergravity side, we note that, both the 
$p+1$d SYM $\rightarrow$ $1+1$d SYM transition
and the Matrix black hole $\rightarrow$ Matrix string 
transition are correspondence regimes where
the geometry, that of a near extremal fundamental
string and that of a black hole respectively, has curvature at the
horizon of order of the string scale~\cite{CORR1}. 
On the other hand,
the BFKS transition is that of longitudinal
localization of the supergravity solution~\cite{HORMART}.

We can now understand the observation of Susskind from the 
phase diagram of Figure~\ref{newfig3}.
From the interacting $p+1$d SYM side, 
one can consider the effective box sizes (\ie\ the critical thermal
wavelengths) as one approaches the various transitions.
The effective box size is defined by
$T_c\Sigma_{\rm eff}\sim 1$, where as usual
the temperature is determined from $T\sim E/S$.
This yields for the $N\sim S$ transition
\bb\label{eff1}
\Sigma_{\rm eff}\sim \Sigma_p \lk(N g_s^2\re)^{2/(9-p)}\ ,
\ee
and for the Matrix String/$p+1$d SYM transition
\bb\label{eff2}
\Sigma_{\rm eff}\sim \Sigma_i \sqrt{N}=\Sigma_p g_s \sqrt{N}\ .
\ee
The bound of Susskind (equation (3.5) of \cite{SUSSGW})
is simply that, starting with the $p+1$d SYM phase at
high temperature, one sees a transition to the Matrix black hole
phase as the temperature is lowered only if
(in the IIA variables)
\bb
g_s>N^{-1/2}\ .
\ee
This is clear from Figure~\ref{newfig3}.

Finally, we note that we assumed above that there exists a well
defined matrix string
description for $N<S$. In this
regime, the thermal wavelength on the Matrix string is smaller than the 
UV cutoff imposed by the discretized nature of the matrices.
Our procedure may be equivalent to an analytical continuation of
the matrix string phase into a regime where the description may not be fully
justified; this is in the same spirit as the extension of the Van der Waals
equation of state into the gas-liquid coexistence region,
which one uses to identify the
emergence of the liquid phase\cite{RIEF}. 
For small enough coupling $g_s$, we expect 
the Matrix string to evaporate
into a perturbative SYM gas, as shown on 
Figure~\ref{newfig3}. Furthermore,
the regime $N<S$ is similar 
to the Hagedorn regime~\cite{HAG,ATTWIT},
in that the temperature remains constant as the system absorbs heat.
We speculate that the $N<S$ regime of the Matrix string near the triple point
is characterized by a coexistent phase of a string with SYM vapor. We defer
a detailed analysis of this issue to future work.

As a unifying probe for all the transitions, we observe that the `mass
per unit charge' $q$ defined in~\pref{rwrtm}
scales on the various transition curves as

\vspace{12pt}
\begin{tabular}{lcl}
Matrix String-$p+1$d SYM Transition         
	& $\rightarrow$ & $q^{-1}\sim g_{\rm eff} \lstr$ \\
Matrix String-Coexistence Phase Transition 
	& $\rightarrow$ & $q^{-1}\sim \lstr$ \\
Matrix String-Black Hole Transition        
	& $\rightarrow$ & $q^{-1}\sim g_{\rm eff}^2 \lstr$ \\
Black Hole-$p+1$d SYM Transition            
	& $\rightarrow$ & $q^{-1}\sim g_{\rm eff}^{2/(9-p)} \lstr$ 
\end{tabular}
\vspace{12pt}

\noindent
with the effective coupling
\bb
g_{\rm eff}^2\equiv g_s^2 N\ .
\ee

From the point of view of the DLCQ string theory 
characterized by the parameters
$g_s$, $\lstr$ and $N$, this scaling on the transition curves
is a non-trivial signature of a unifying framework
underlying the physics of criticality of the theory. 
Note also that the $g_s^2 N$ combination 
is {\em not} the 't Hooft coupling of the
Matrix SYM description, equation~\pref{maindict};
recall that $g_s$ is a modulus
of the torus compactification.

From the point of view of field theory, the various transitions
that we have identified are predictions about
the thermodynamics of 4+1d, 5+1d SYM on the torus well into
non-perturbative field theory regimes.

\section{Comments about charged phases}

In this section, we will focus on the triple point, 
where the correspondence and localization effects coincide, 
and present general considerations of relevance
to singly and doubly charged black holes. 
We will see that, at the critical point,
charged black holes are characterized by $N/S>1$. 
For the singly charged case,
by making use of 't Hooft holonomies on the torus,  we will sketch
a simple dynamical mechanism by which the 
system can cluster its SYM excitations
at the triple point so as to account for the ratio $N/S$.

Let us begin by rephrasing part of our previous analysis
in a slightly different language, using the IMF formalism. 
Consider a black hole 
in $D$ dimensions arising from M-theory on $T^{p+1}$, and define $D+p=10$.
Denote the size of the M-theory circle
by $R$, and suppose the other circles have the characteristic sizes $R_i$
as in Section 2.
The horizon radius $r_0$ of a black hole of mass $M$ is
\bb \label{bhmass}
  r_0^{D-3}=\frac{\lpl^9}{RR_1\cdots R_{p}}M\sim 
	\Bigl(\frac{\lpl^3}{R}\Bigr)^{(D-3)/2}
\ee
at the correspondence point where the horizon size is the string
scale $\lstr^2\sim \lpl^3/R$.  To reach the correspondence
point for a given mass black hole, one tunes the string 
coupling $\gstr\sim (\lpl/R)^{3/2}$ while holding
the size of the compactification torus (other
than the M-theory circle) fixed in string units.
Therefore let all the $R_i=\lambda\lstr$; one finds
\bb \label{corrmass}
  M\lstr\sim \lam^{d-1}\gstr^{-2}\ .
\ee
At this point, the string entropy 
\bb \label{strent}
  S_\str\sim \sqrt N_\osc
\ee
is of the same order of magnitude as the black hole entropy
($G_\D$ is the $D$-dimensional Newton constant)
\bb\label{bhent}
  S_\bh\sim r_0^{D-2}G_\D^{-1}\ .
\ee

To work at the triple point,
we demand that the string coupling is tuned so that
the horizon size is of order the string scale,
as dictated by the correspondence principle, and that
the black hole is placed in a small box and boosted
to the black hole/black string transition 
\cite{BFKS1,BFKS2,BFKS3,HORMART,LAFLAMME1,LAFLAMME2},
so that it can be described by matrix theory as a D-brane fluid.
At the latter point, the entropy of an uncharged black hole
is related to the momentum $P=N/R$ of the boosted hole 
by $S_\bh\sim N=RP$.  In terms of general relativity,
the effect of the boost is to expand the proper size of the
box near the black hole so that it `just fits inside'.
For a small box and a large black hole, the system is highly
boosted at the string/hole transition.
In the IMF, the weak-coupling string 
that (according to the correspondence principle)
approximates the black hole, has an energy 
\bb\label{elc}
  E_\lc\sim \Sigma T^2 \sim \frac{1}{P\lstr^2}N_\osc\ ,
\ee
where $\Sigma$ is the length of the string, $T$ is its temperature,
$N_\osc$ is the oscillator excitation number, and
$P=N/R$ is the longitudinal momentum.  From this we
find the temperature is given by
\bb\label{strtemp}
  T\sim \frac{R N_\osc^{1/2}}{N\lstr^2}\ .
\ee
On the other hand, one expects the black hole to emit
Hawking radiation at temperature\footnote{We remind the reader
the IMF kinematics $E_{LC}\sim M e^{-\alpha}$ and
$P\sim M e^\alpha$.}
\bb\label{thawk}
  T_H\sim \frac{e^{-\alpha}}{r_0}\sim \frac{R}{r_0^2}\ ,
\ee
where $\alpha$ is the rapidity of the boost needed to 
fit the hole in the longitudinal box.  The temperature 
of emitted quanta must be the same as 
the temperature of the gas on the string; equating the two
expressions \pref{strtemp} and \pref{thawk}, we find
\bb\label{tempcompare}
  \frac{N_\osc^{1/2}}{N}\sim\frac{\lstr^2}{r_0^2}\ .
\ee
Now the boost quantum $N$ is the entropy of the black hole,
as is the square root of the oscillator number; the
left-hand side is of order unity, and therefore
the black hole is of order the string size. 
This is a rephrasing of the observation 
\pref{corr} of Section 3, at the triple point.
Moving away from this point by boosting to $N>S$ 
is a canonical operation;
$S\sim \Sigma T$ for a string, and $\Sigma\propto N$
while $T\propto 1/N$.  Note that the number of matrix partons
in a thermal wavelength of the matrix string at the
correspondence point is always $N_{\rm cl}\sim N/S$,
in agreement with the proposal of \cite{LIMART}.

For a singly-charged hole, the mass, charge and entropy
are given by
\bbb
  M &\sim &~  G_D^{-1}r_0^{D-3}(\ch^2\gamma+\coeff{1}{D-3}) \nonumber \\
  Q &\sim &~  G_D^{-1}r_0^{D-3} \sh\gamma\ch\gamma \label{chbh}\\
  S &\sim &~  G_D^{-1}r_0^{D-2} \ch\gamma\ \hfill \nonumber
\eee
in terms of the `charge rapidity' $\gamma$.
The Hawking temperature in the boosted frame is
\bb\label{chthawk}
  T_H\sim \frac{e^{-\alpha}}{r_0\;\ch\,\gamma}\ ;
\ee
again equating this temperature with the string temperature \pref{strtemp}
and setting the horizon size equal to the string scale 
correctly yields
\bb\label{chent}
  S_\bh\sim N_\osc^{1/2}\sim \frac{N}{\ch\,\gamma}\ .
\ee

This implies that, for the charged case, even at the BFKS point, $N/S >1$.
This ratio was interpreted in~\cite{LIMART} as the size of clusters of
partons making up the black hole phase. Therefore, the charged black hole
is to be described at the BFKS point as a gas 
of clusters of size $\ch \gamma$.

We can see a mechanism for this dynamics with the following argument. 
Consider the Matrix string limit of Matrix theory on $T^2$.
Matrix string transverse excitations are stored 
in the scalars of the $2+1$ SYM,
on the diagonal of the matrices, with the eigenvalues sewn by
boundary conditions involving the shift operator~\cite{DVV}.
Off-diagonal constant modes describe the effect of the
W bosons stretched between the strands of the
string; their one-loop fluctuations give 
the effective gravitational interaction
of the bits of the Matrix string. Given a holonomy describing
a singly charged Matrix string phase, one can ask what is the interaction
potential between two points on the strands, or between two matrix elements 
on the D-string. Fixing a winding charge $Q$, 
a magnetic field must be turned on
\bb
B_{12}=\frac{R}{N} Q\ .
\ee
This induces a holonomy on the 2-torus, with 't Hooft style boundary
conditions~\cite{THOOFT}. Extending these conditions onto the
scalars $X^i$, we get
\bbb
  X^i&=&~V^k X^i V^{-k}\label{thooftbc}\\
  X^i(\sigma+\Sigma)&=&~ U^l X^i(\sigma) U^{-l}\hfill \ ,\nonumber
\eee
with $U_{ij}=\delta_{i,j-1}$, 
$V_{ij}= diag\lk[\exp(2\pi i(m-1)/N)\re]$, $m=1,...,N$;
and $k l=\alpha' Q/R_i=\lstr Q$. 
This configuration describes a system consisting of $l$ strings,
and requires the vanishing of some of the off-diagonal 
elements between the matrix strands. This latter effect reduces the 
strength of the interaction of the strings
when the off-diagonal modes are integrated, by a $Q$ and $l$ dependent factor. 
Modeling the system through
the interaction of the zero modes, and taking into account the 
multiplicity factor in the interaction 
resulting from the boundary conditions \pref{thooftbc},
we find an energy for the gas as a function of $l$
\bb
E(l)\sim \frac{N}{l R} v^2- K l \lk(\frac{N}{l}\re)^2 
\frac{Q \lstr}{N} \frac{G_D}{R^3} \frac{v^4}{r^{7-d}}\ ,
\ee
with $K$ a numerical coefficient independent of $N$ and $l$. 
Applying the uncertainty principle and the virial
theorem~\cite{HORMART} at the correspondence point $r_0\sim \lstr$
then yields the scaling
\bb
\frac{N}{l} \sim \ch \gamma\ ,
\ee
in agreement with~\pref{chent}; \ie,  it is energetically favorable for the
system to settle into a phase of clusters of matrices 
of size $N/S\sim \ch \gamma$.

Finally, a doubly-charged hole is obtained in the correspondence
principle from a string carrying both winding and momentum
in compact directions.  The IMF energy and the worldsheet
temperatures of left- and right-movers are
\bbb
  E_\lc&\sim &~ \Sigma(T_L^2+T_R^2)\sim\frac{N^\osc_L+N^\osc_R}{P\lstr^2}
\nonumber \\
  T_{L,R}&\sim &~ \frac{R N_{L,R}^{1/2}}{N \lstr^2}\hfill\ . 
\label{dchtemps}
\eee
At the string-hole transition, one has
\bbb
  P&\sim &~ G_\D^{-1}r_0^{D-3}(\ch^2\gamma_w+\ch^2\gamma_p)e^\alpha
\nonumber \\
  S&\sim &~ G_\D^{-1}r_0^{D-2}\ch\,\gamma_w~\ch\,\gamma_p\hfill\ , 
\label{strhole} 
\eee
from which one finds the relation
\bb\label{NSrel}
  N\sim S\;\frac{\ch^2\gamma_w+\ch^2\gamma_p}{\ch\,\gamma_w~\ch\,\gamma_p}
\ee
between the longitudinal boost quantum and the entropy.  Note that 
$S\sim N_{\osc,L}^{1/2}+N_{\osc,R}^{1/2}\propto \ch\gamma_w\ch\gamma_p$,
while the Hawking temperature must be 
$T_H=\frac{2T_LT_R}{T_L+T_R}\propto (\ch\gamma_w\ch\gamma_p)^{-1}$.
A consistent assignment of worldsheet oscillator numbers
with respect to these quantities is
\bb\label{oscnums}
  N_{L,R}\sim \lk(G_D^{-1}r_0^{D-2}\ch(\gamma_w\pm\gamma_p)\re)^2\ ,
\ee
so that the oscillator entropy agrees with the
black hole entropy \pref{strhole}, and the Hawking temperature 
determined from \pref{dchtemps} is of the right order
\footnote{These expressions
are somewhat different than those in \cite{CORR1}; the point is that
one is free to adjust the smaller of the two temperatures $T_{L,R}$
without appreciably affecting the entropy or the charges.
Our choice is compatible with the BPS limit 
$r_0\rightarrow 0$ with $\gamma_w\sim\gamma_p$ and the charges fixed,
whereas the one in~\cite{CORR1} does not give vanishing $T_R$.}. We note that
we have again $N/S>1$ at the triple point due to the presence of KK charges.

Thus, in general it is necessary that the matrix black hole
consist of coherent clusters of matrix partons
even at the BFKS transition; and that, at the correspondence point,
the number of partons in a cluster is the same as the number
of strands of the matrix string lying within a thermal wavelength.
Both of these facts support the analysis of~\cite{LIMART}.

\section{The interacting Matrix string}

In this section, we come back to the neutral case, away from the triple point,
attempting to probe the dynamics in greater detail
from the Matrix String side. In~\cite{LIMART}, it was argued that
the Matrix black hole phase, as a SYM field configuration,
can be thought of as a gas of $S$ clusters of
D0 branes, the zero modes of the SYM, 
each cluster consisting of $N/S$ partons. The system
is self-interacting through the $v^4/r^7$ interaction, or its smeared
form on the torus.

This phase of clustered D0 branes may be an effective description, \ie\
thermodynamically strongly correlated regions of a metastable state; 
or more optimistically, it might 
be a microscopic description associated with formation of bound
states like in BCS theory. We will try here to investigate the Matrix
string dynamics so as to reveal the signature of the clusters as we
approach the correspondence curve. The aim is to identify a possible
dynamical mechanism for black hole formation, 
and determine the correspondence curve from such
a microscopic consideration.

In Section 5.1, we derive 
the potential between two points on the Matrix string;
in view of the Matrix conjecture, 
we can do this by expanding the DBI action
of a D-string in the background of a D-string. We then evaluate the
expectation value of this potential in the free string ensemble at 
fixed temperature. In Section 5.2, we analyze the characteristic features of
the potential, particularly noting the bump for $p=4,5$ that we alluded to
in the Introduction. In Section 5.3, we comment on the dynamics implied
by the potential, particularly in regard to the phase diagram derived 
earlier.

\subsection{The potential}

In this section, we derive the potential between two points on the
Matrix string by expanding 
the DBI action for a D-string probe in the background
geometry of a D-string.
We then check the validity of the DBI expansion, and
evaluate the expectation value of the potential in the thermal ensemble 
of highly excited Matrix strings. The details of the finite temperature
field theory calculations are collected in Appendix A.

A IIA string in Matrix string theory is constructed in a sector of
field configurations described by diagonal matrices, 
with a holonomy in $Z_N$;
a nonlocal gauge transformation converts this
into 't Hooft-like twisted boundary conditions on the
transverse excitations of the eigenvalues of the matrices, as described
in detail in \cite{DVV,BSMAT,WYNMAT,MOTL}.
The conclusion is that the eigenvalues are sewn
together into a long string, and a IIA string emerges as an object looking
much like a coil or `slinky' 
wrapped on $\Sigma_p$. The self-interactions of this
string are described by integrating out off-diagonal modes between the
well-separated strands. Alternatively, making use of the Matrix
conjecture, this effective action can be obtained from supergavity, by
expanding the Born-Infeld action of 
a D-string in the background of a D-string.
\footnote{Note that, for D-string strands closer to each other than
the Plank scale, the W bosons cannot be integrated out of the problem;
the physics is described by the full non-abelian degrees of freedom. 
We are assuming here that this `UV' physics does not effect the analysis 
done at a larger length scale.}
We will follow this prescription to calculate 
the gravitational self-interaction
potential between two points on a highly excited Matrix string.

The Born-Infeld action for N D-strings is given by~\cite{POLCHTASI}
\bb \label{DBI}
S=-\frac{1}{2 \bar{\alpha}' \bar{g}_s}
 \lk[ \int d^2 \sigma e^{-\phi} \mbox{Tr} \mbox{Det}^{1/2}
\lk( G_{ab} + B_{ab}+2 \pi \alpha' F_{ab} \re) - N \int C_{RR}^{(2)}
\re] ,
\ee
where we have assumed commuting matrices so that there is no ambiguity in
matrix orderings in the expansion, and $\bar{g}_s$ is the dilaton vev 
at infinity. We choose $\sigma^1$ to have radius $\Sigma_p$,
turn off gauge and NS-NS fluxes,
\bb
B_{ab}=F_{ab}=0\ ,
\ee
and choose the static gauge
\bbb
X^0&=&~\sigma^0 \mbox{1}\ ,\\
X^1&=&~\sigma^1 \mbox{1}\ .\nonumber
\eee
A single D-string background in the string frame is given by~\cite{SL2Z}
\bbb
ds_{10}^2&=&~h^{-1/2} (-dt^2+dx^2)+h^{1/2} (d\vec{x})^2 \nonumber\\
\medskip
e^\phi&=&~h^{1/2}\hfill \\
C_{01}&=&~h^{-1}\hfill\nonumber\\
h&=&~\lk( \frac{r_0}{r} \re)^{7-p} \ ,\nonumber
\eee
(recall $p$ is the torus dimension)
and we define $D\equiv 7-p$. 
The string is taken to have no polarizations
on the torus, nor any KK charges.
Here, we have followed the prescription in \cite{BECKERS,KRAUS}, 
where we T-dualized the D-string 
solution to a D0 brane, lifted to 11 dimensions, 
compactified on a lightlike direction,
and T-dualized the solution to the one above.
The only change is in replacing $1+(r_0/r)^{7-p}\rightarrow
(r_0/r)^{7-p}$. By Gauss' law, 
\bbb
r_0^{D}&=&~c_D \frac{g_s^2}{R_+^2} \lstr^{9-p}\ ,\\
c_D&\equiv&~ \frac{(2\pi)^7}{2\pi^{D/2}} \Gamma(D/2)\ ,\nonumber
\eee
where we have made use of the needed dualities to express things in our
IIA description.
Putting in the background, we have
\bbb\label{bistr}
S&=&~-\frac{1}{\alpha'}\int^{\Sigma_p N}\ 
h^{-1}\lk(1+hK+h^2V\re)^{1/2}-h^{-1}\\
K&\equiv&~ {X'}^2-{\dot{X}}^2=4\del_+ X.\del_- X\nonumber\\
\medskip
V&\equiv&~ 4((\del_+X.\del_-X)^2-(\del_+X)^2 (\del_-X)^2)\ .\nonumber
\eee
We note that, as the limit of the action indicates, we have made use of
the $Z_N$ holonomy that sews the rings of the slinky together.
Expanding the square root yields the Hamiltonian
\bb \label{hamiltonian}
H=\int^{\Sigma_p N} \frac{1}{2 \alpha'} ({\dot{X}}^2+{X'}^2)
+c_D'\frac{g_s^2 \lstr^{7-p}}{R_+^2 r^{7-p}}
\lk\{(\del_+X)^2 (\del_-X)^2-\lk[ (\del_+X)^2+(\del_-X)^2\re] 
(\del_+X.\del_-X) \re\}\ .
\ee
Let us check the validity of the DBI expansion we have performed. 
We would like to study dynamics of the string squeezed 
at most up to the string scale, the
correspondence point; setting $r\sim \lstr$ in $h$, we get
\bb \label{hh}
h\sim \lk(\frac{g_s \lstr}{R_+}\re)^2\ .
\ee
From elementary string dynamics 
(equations \pref{strent},\pref{elc}), we have
\bb\label{kkk}
\vev{K}\sim \lk(\frac{R_+ S}{\lstr N}\re)^2\ ,
\ee
where brackets indicate thermal averaging at fixed entropy $S$. It can be
shown from the results of the next section that
\bb\label{vvv}
\vev{V}_{max}\sim \vev{K}^2\ ,
\ee
and that $\vev{V}$ will have a definite maximum for all $p$. 
We now see from \pref{bistr},\pref{hh},\pref{kkk}, and \pref{vvv},
that our DBI expansion is a perturbative expansion in
\bb
\varepsilon= g_s \frac{S}{N}\ .
\ee
We then need
\bb
S \ll N g_s^{-1}\ .
\ee
A glance at Figure~\ref{newfig3} reveals that we are well within the region
of interest.

The potential in this expression is the interaction energy 
between a D-string probe and a D-string source.  Using the residual
Galilean symmetry in the DLCQ, and assuming string thermal wavelengths
$> \Sigma_p$ (the `slinky regime'), we deduce that the potential between
two points on the Matrix strings denoted by the labels $1$ and $2$ is
\bb \label{slinky}
V_{12} = {K}_D \frac{g_s^2 \lstr^{5-p}}{R_+} \frac{
\lk\{(\del_+X_r)^2 (\del_-X_r)^2-\lk[ (\del_+X_r)^2+(\del_-X_r)^2\re] 
(\del_+X_r.\del_-X_r)
\re\}}{\lk(X_r^2\re)^{D/2}}\ ,
\ee
where
\bb
X_r \equiv X_2-X_1 .
\ee
and $K_D$ is a horrific numerical coefficient we are
not interested in.

Ideally, one should self-consistently determine the shape distribution
of the string in the presence of this self-interaction;
however, this is rather too complicated to actually carry out.
To first order in small $g_s$, 
the effect of the potential is to weigh different regions of the
energy shell in phase space~\cite{GOLDENFELD} by a factor derived from 
its expectation value in the free string ensemble. 
We will discuss the dynamics in the presence of the potential
in somewhat more detail below.
For now, in light of this weak-coupling approximation scheme,
we would like to calculate the expectation value of the potential
in a thermodynamic ensemble consisting of a highly excited 
{\it free} string with fixed entropy $S$.
From the Matrix string theory point of view, this is essentially a 
problem in finite
temperature field theory, where we will deal with a two dimensional 
Bose gas (ignoring supersymmetry; the fermion contribution is similar) 
on a torus with sides $\Sigma_p N$ and $\beta=1/T$,
$\beta$ being the period of the Euclidean time. Using Wick contractions,
we can then express the potential in terms of the free Green's functions;
we defer the details to Appendix A.
We get
\bb \label{thepotential}
V_{12} = \alpha_D g_s^2 \frac{\lstr^{5-p}}{R_+}
\frac{K^{zz}_{12} K^{\bar{z}\bar{z}}_{12}}{\lk( -K_{12} \re)^{D/2}} ,
\ee
where $\alpha_D$ is a dimension dependent numerical coefficient,
\bb
K_{12}\equiv K_\Delta\equiv -\alpha' \lk<X_1 X_2\re>
\equiv -\alpha' G_{12}\ ,
\ee
is the Green's function of the two dimensional Laplacian
on the torus, and $K^{zz}_{12}$ is its double derivative with respect
to the $z$ complex coordinate of the Riemann surface representing
the Euclideanized world-sheet.
We refer the reader to Appendix A for the derivation of this equation.

\subsection{The thermal free string}

The thermodynamic properties of the matrix string at
inverse temperature $\beta$ are determined by the Green's function
of the Laplacian on the worldsheet torus of sides
$(\Sigma,\beta)$, where $\Sigma\equiv \Sigma_p N$.
It is known from CFT on the torus 
that this is given by\cite{DIFRANCESCO,KIRITSIS}
\bb
G_{12}=-\frac{1}{2 \pi} \ln \lk| 
\frac{\theta_1\lk(\frac{z}{\Sigma} | \tau \re)}
{\theta'_1\lk( 0| \tau \re)} \re| +\frac{1}{2 \tau_2} 
\lk( \mbox{Im} \frac{z}{\Sigma}
\re)^2 ,
\ee
where
\bb
\tau\equiv i\frac{\beta}{\Sigma}\equiv i \tau_2= \frac{i}{S} .
\ee
Here $\beta$ can be obtained from the 
free string thermodynamics of equation~\pref{strtemp}.
All correlators and their derivatives must eventually 
be evaluated on a time slice
corresponding to the real axis in the $z$ plane.

Divergences will be seen
in correlators due to infinite zero point energies.
The conventional approach
is to introduce a normal ordering scheme giving the vacuum zero
expectation value in such situations, \ie\ throwing away disconnected
vacuum bubbles. 
In our case, the string has a classical background due
to its thermal excitation.
To renormalize finite $T$ correlators, 
we subtract the zero temperature limit from each propagator. 
This corresponds to
\bb
\ave{f(X)} \sim f(\coeff\delta{\delta J}) \ln Z_T[J]
\rightarrow f(\coeff\delta{\delta J}) \ln Z_T[J] - 
f(\coeff\delta{\delta J}) \ln Z_{T=0}[J]=
f(\coeff\delta{\delta J}) \ln \Bigl(\frac{Z_T[J]}{Z_{T=0}[J]}\Bigr) .
\ee
From the expression for $Z[J]$, we see that this amounts to correcting the
Green's functions as
\bb
K \rightarrow K_T - K_{T=0} .
\ee
This subtraction removes the divergent zero-point fluctuations
of nearby points on the string, while leaving the effects
due to thermal fluctuations.

\begin{figure}[t]
\epsfxsize=7cm \centerline{\leavevmode \epsfbox{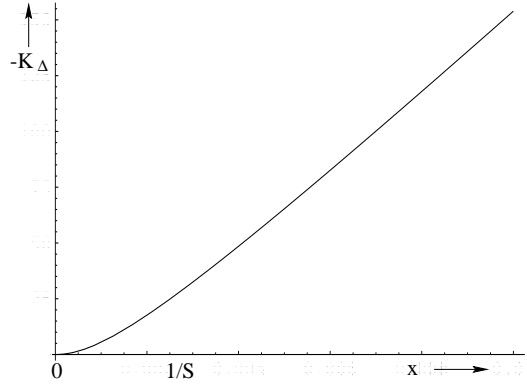}}
\caption{\sl $-K_\Delta$ as a function of 
the string separation parameter $x$;
we see the change of scaling from $x^2$ to $x$.}
\label{f2}
\end{figure}

Defining
\bb
x\equiv \frac{z}{\Sigma}\ ,
\ee
we then have, subtracting the zero temperature part,
\bb
K_\Delta\rightarrow \alpha' \lk(\frac{1}{4 \pi} 
\ln g \bar{g} +\frac{1}{8 \tau_2} \lk( x-\bar{x} \re)^2 \re) ,
\ee
with
\bb
\ln g=\sum_{n=1} \ln \lk( \frac{1- 2 q^n \cos (2 \pi x) + q^{2 n}}
{\lk( 1-q^n \re)^2} \re) .
\ee
We can now make use of $\tau_2\ll 1$ for $S\gg 1$, to write
this sum as an integral
\bb
\ln g=\frac{1}{2 \pi \tau_2} \int_0^{e^{-2 \pi \tau_2}} \frac{dv}{v} \ \ 
\ln \lk( \frac{1-2v \cos (2 \pi x) + v^2}{\lk( 1-v \re)^2} \re) .
\ee
This integral can be evaluated to yield
\bb\label{Kfinal}
\ln g =\frac{1}{2 \pi \tau_2} \lk( 2 Li_2 e^{-2 \pi \tau_2}
-Li_2 e^{-2 \pi \tau_2 + 2 \pi x i}
-Li_2 e^{-2 \pi \tau_2 - 2 \pi x i} \re) ,
\ee
where $Li_2$ is the PolyLog function of base 2, related to the Lerch $\Phi$
function~\cite{MATH}. We then have
\bb\label{kdelta}
K_\Delta=\frac{\alpha'}{2\pi} \ln g\ ,
\ee
with $x$ here being real, and
representing the equal-time separation between
two points on the string, $x=x_1-x_2$, 
as a fraction of the total length $\Sigma$ ($0<x<1$).
The asymptotics are
\bb
K_\Delta \simeq \lk[ 
\begin{array}{cl}
\alpha' S x & \mbox{     for     } \tau_2 \ll 1\\
\frac{\alpha'}{\pi} S^2 x^2 & \mbox{     for     } 2 \pi x \ll 1
\end{array}
\re . ,
\ee
The first line is a well-known 
result of Mitchell and Turok~\cite{MITCHTUR} calculated 
originally using the microcanonical
ensemble. It shows random walk scaling $\sqrt{\ave{R^2}}\sim N_\osc^{1/4}
x^{1/2}$. The second line is new 
and valid for small separations on the string;
it is the statement that within 
the thermal wavelength $\beta$ of the
the excited string, the string is stretched, scaling as 
$\sqrt{\ave{R^2}}\sim N_\osc^{1/2} x$. 
This is intuitively expected, as regions
on the string within the typical thermal 
wavelength will be strongly correlated
in the thermodynamic sense. This change in the scaling is crucial to
what we will soon see in the behavior of the potential between strands.
$-K_\Delta$ is plotted in Figure~\ref{f2}.

Next, consider the derivatives of the correlators, 
evaluated on the real axis. 
We have
\bb
\del_x K_\Delta = \del_{\bar{x}} K_\Delta =
\frac{-i \alpha'}{2 (2 \pi)^2 \tau_2} \ln 
\lk( \frac{1-e^{-2\pi \tau_2 -2 \pi i x}}
{1-e^{-2\pi \tau_2 +2 \pi i x}} \re) .
\ee
We also have
\bb
\del_x \del_{\bar{x}} K_\Delta =\frac{-\alpha'}{4 \tau_2} ,
\ee
or
\bb
K^{z \bar{z}}_\Delta\rightarrow 0 ,
\ee
since we subtract the zero temperature result.
The most relevant term is
\bb
\del_x^2 K_\Delta= \del_{\bar{x}}^2 K_\Delta =
-\frac{\alpha'}{\tau_2} \frac{1-e^{2 \pi \tau_2} \cos (2 \pi x)}
{e^{4 \pi \tau_2} - 2 e^{2 \pi \tau_2} \cos(2 \pi x) +1}
+\frac{\alpha'}{4 \tau_2} ,
\ee
or
\bb \label{Kzzfinal}
\Sigma^2 K^{zz}_\Delta=\Sigma^2 K^{\bar{z} \bar{z}}_\Delta \rightarrow
-\frac{\alpha'}{\tau_2} \frac{1-e^{2 \pi \tau_2} \cos (2 \pi x)}
{e^{4 \pi \tau_2} - 2 e^{2 \pi \tau_2} \cos(2 \pi x) +1}
-\frac{\alpha'}{\tau_2} \frac{1}{e^{2 \pi \tau_2}-1} 
\ee
(again we subtract the zero temperature part).

\begin{figure}[t]
\epsfxsize=7cm \centerline{\leavevmode \epsfbox{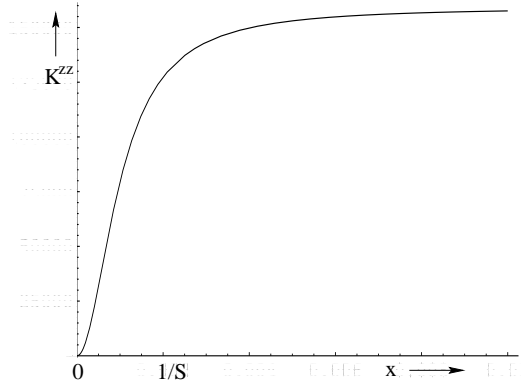}}
\caption{\sl $K^{zz}_\Delta$ as a function 
of the string separation parameter $x$;
we see the flattening of the correlation at large $x$. For small $x$, small
relative stretching or motion is implied; for larger $x$, the flattening
indicates a constant correlation in the relative stretching of the string.}
\label{barkzz}
\end{figure}

This yields the asymptotics
\bb
(N\Sigma_p)^2 K^{zz}_\Delta\simeq \lk[ 
\begin{array}{cl}
\alpha' S^2 +\frac{\pi}{12} (5+\cos(2\pi x)) (\csc(\pi x))^2 +O(\tau_2^2)
\rightarrow \alpha' S^2 & \mbox{     for     } \tau_2 \ll 1\\
\alpha' S^4 x^2 & \mbox{     for     } 2 \pi x \ll 1
\end{array}
\re . .
\ee
$K^{zz}_\Delta$ is plotted in Figure~\ref{barkzz} as a function of $x$.

\subsection{The bump potential}

We now put together equations~\pref{kdelta} and~\pref{Kzzfinal}
in the potential of~\pref{thepotential} to get
the asymptotics
\bb
V_{12}\simeq g_s^2\frac{R_+^3}{{\alpha'}^3 N^4} \lk[
\begin{array}{cl}
S^{(8-D)/2} x^{-D/2} & \mbox{     for     } \tau_2 \ll 1\\
S^{8-D} x^{4-D} & \mbox{     for     } 2 \pi x \ll 1
\end{array}
\re . .
\ee
For $p>3$, $V_{12}\rightarrow 0$ as $x\rightarrow 0$; 
at larger $x$, it decays as $x^{-D/2}$. 
At the thermal wavelength $x\sim 1/S$, both expressions give
\bb
V_{max}\simeq g_s^2 \frac{R_+^3}{{\alpha'}^3 N^4} S^4 .
\ee
Note that in this expression the dimension dependence in the power of $S$
conspires to vanish.
For $p=3$, $V_{12}\sim S^4$ for $x\rightarrow 0$, while for $p<3$,
$V_{12}\rightarrow \infty$ for $x\rightarrow 0$;
in both of these latter cases,
the potential decays as $x^{-D/2}$ for larger $x$.

\begin{figure}[t]
\epsfxsize=7cm \centerline{\leavevmode \epsfbox{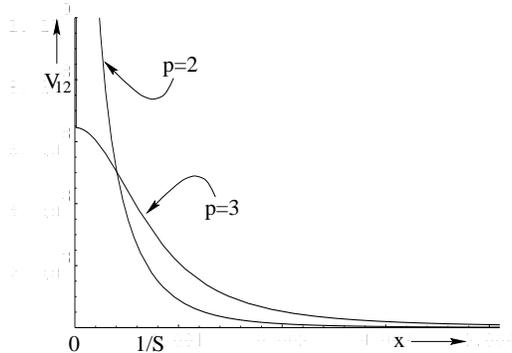}}
\caption{\sl The potential as a function of $x$ 
for dimensions $p=3$ and $p=2$.}
\label{pot2}
\end{figure}

The conclusion can be summarized as follows. 
For $p=4$ and $p=5$, there exists a
bump in the potential of height proportional to $S^4$ at the
thermal wavelength on the string; for $p=3$, the bump smoothes to a flat
configuration where the difference between the potential at the thermal
wavelength separation and at $x=0$ is of order unity. Finally, for $p<3$,
the bump disappears altogether and the potential blows up at the origin
signaling the breakdown of the description.
This potential is plotted in various cases 
in Figure~\ref{potl} of the introduction and Figure~\ref{pot2}.

The presence or absence of the bump is a result of two competing
effects: First of all, the increasingly singular short-distance
behavior of the Coulomb potential \pref{slinky} 
with increasing dimension $D$;
and secondly, the strong correlation of neighboring points
on the string, which makes $(\d X_1-\d X_2)$
decrease as the separation along the string decreases
(inside a thermal wavelength).

We observe that:
\begin{itemize}
\item
The bump occurs at separations of $1/S$ of a fraction of the 
whole length of the string; in the matrix language, this corresponds to
a bump about matrices of size $N/S$.
\item
The presence or absence of the bump 
as a function of the number of non-compact
space dimensions correlates with the observations of~\cite{CORR2}, given
that in the DLCQ, the light-like direction reduces the number of 
non-compact dimensions by one.
\item
As described in~\cite{LIMART}, a matrix black hole
can be described by SYM excitations clustered within matrices of
size $N/S$, the location of the bump. Furthermore, we will shortly
reproduce, from scaling arguments regarding the dynamics of this
potential, the two correspondence lines determined from thermodynamic
considerations above.
\end{itemize}

\noindent
We then conclude that we have identified the characteristic signature of
black hole formation in the Matrix SYM.

\subsection{Dynamical issues and criticality}

The dynamics of this potential near 
a phase transition point is certainly complicated. 
Intuitively, we expect that as we approach a critical point, instabilities
develop, an order parameter fluctuates violently, perhaps 
related to some measure of the $Z_N$ symmetry;
it is reasonable to expect the characteristic feature of the potential,
the confining bump, plays a crucial role in the dynamics of the 
emerging phase.
Deferring a more detailed analysis of these issues to
the future, let us try to extract from these results the scaling of the
correspondence curves.

First let us motivate the use of 
the expectation value of the potential
in the free string ensemble.  We indicated earlier that
this quantity is qualitatively related to the effect of the
interactions, assuming they are weak enough, on the
energy shell in phase space covered by the free string.
The partition function becomes, schematically
\bb
Z \sim \mbox{Tr } e^{H_0+V} \sim e^{\lk< V\re>_0} \mbox{Tr } e^{H_0}\ ,
\ee
so that phase space is weighed by an additional factor related to
the expectation value of the potential in the free ensemble 
$\lk< V\re>_0$. This is also similar to 
the RG procedure applied to the 2d Ising model, where the 
context and interpretation 
is slightly different~\cite{GOLDENFELD}.

Using equation~\pref{thepotential}, 
the potential energy content of the matrix
string is given by
\bb\label{potwind}
V\sim \int_0^{N\Sigma_p} d\sigma_-\ 
\lk( S^4 g_s^2 \frac{R_+^3}{{\alpha'}^3 N^4}\re) \lk( N\Sigma_p \re) v_{12}\ ,
\ee
where we have integrated over one of the two intergals of 
the translationally invariant two-body potential, and scaled $v_{12}$
such that its maximum is of order 1, independent of any state variables;
however, the shape of $v_{12}$ still depends on $N$ and $S$. This
expression represents the interaction energy between two points on the
coiled matrix string at fixed separation $\sigma_-$. 
From the point of view of Matrix theory physics, the string's fundamental
dynamical degrees of freedom are the windings on the coil; we expect 
a transition in the dynamics of the object 
when there is a competition between forces on an individual winding.
In the present case, the two forces are nearest neighbor elastic
interaction and the gravitational interaction. 
A single string winding being wrapped on $\Sigma_p$ worth of world-sheet,
the maximum potential energy it feels
can be read from equation~\pref{potwind}
\bb\label{vmaxwind}
v_{max}\sim S^4 g_s^2 \frac{R_+^3}{{\alpha'}^3 N^4} N\Sigma_p^2 \ ,
\ee
and is due to its interaction with strands a thermal wavelength away. 
Its thermal energy caused by nearest neighbour interactions
is read off equation~\pref{kkk}
\bb\label{kinwind}
\kappa\sim \frac{\lk<K\re>}{\alpha'} \Sigma_p\ .
\ee
The two forces compete when
\bb \label{corrr1}
S\sim \sqrt{N} g_s^{-1}\ .
\ee
At stronger coupling, the forces due to the gravitational interaction
dominate those of the nearest neighbor stretching and
decohere neighboring strands' velocities.  The free string
evaluation of the interaction, equation~\pref{thepotential},
is no longer valid; one expects a phase transition to occur.
Equation~\pref{corrr1}
is our matching result of~\pref{spec1} between the string and $p+1$d
interacting SYM phase. Here, we are assuming an analytical continuation
of the Matrix string phase to the region $N<S$ in the phase diagram;
our suggestion that this region is associated with a coexistence phase
is consistent with this procedure.

To account for the correspondence curve for $N>S$,
we now recall that in the discussion of clustered
D0 branes of~\cite{LIMART}, the virial treatment of the $v^4/r^7$ 
interaction had to be corrected by a factor in order to reproduce
the black hole equation of state; the origin of this correction was
argued to be interaction processes between the clusters involving
the exchange of longitudinal momentum. Under the assumption that these
effects are of the same order 
as zero momentum transfer processes, a correction
factor of $N/S$ was applied. 
Using a chain of dualities, we can quantify the effect of longitudinal
momentum transfer physics by studying the scattering amplitude
in IIB string theory with winding number exchange. We do this
in Appendix B, where we find that, for exchanges of windings up to order
$N/S$, the winding exchange generates an interaction
identical to that of zero longitudinal momentum exchange; for higher winding
exchanges, the interactions are much weaker. These winding modes,
represent the sections of the Matrix string within the thermal wavelength, 
$N/S$ worth of D-string windings. 
Thus we modify the $v_{12}$ potential above by the factor $N/S$,
which accounts in the scaling analysis for the effect
of longitudinal momentum transfer physics in the Matrix string
self-interaction potential. Applying the virial theorem between 
equation~\pref{kinwind} and $N/S$ times equation~\pref{vmaxwind}
yields the Matrix string-Matrix black hole correspondence
point at
\bb \label{corr2}
S\sim g_s^{-2}\ ,
\ee
as needed.

We can now interpret our results as follows. The bump potential accounts
for the matching of the string phase onto {\it both} 
$N<S$ and $N>S$ phases, one involving partons interacting
without longitudinal momentum exchange 
(the Matrix string-($p+1$d) SYM curve in
Figure~\ref{newfig3}), and the other being the 
Matrix black hole phase of parton clusters of
size $N/S>1$ interacting in addition 
by exchange of longitudinal momentum 
(the matrix string/matrix black hole
correspondence curve of Figure~\ref{newfig3}). 
In the latter case, the location
of the confining bump correlates with 
matrices of size $N/S$. In the former
case, the correlations are finer than 
the UV matrix cutoff; a better understanding
of this latter issue obviously 
needs a more quantitative analysis of the $N<S$ 
Matrix string regime. This analysis further substantiates the
identification of the bump potential as the signature of black
hole formation from Matrix SYM, as well as justifying the new
matrix string-$p$ brane transition
microscopically.

\vskip 2cm
{\Large{{\bf Acknowledgments}}}
\vskip 1cm

We are grateful to H. Awata for discussions. V.S. is very grateful to
S. Coppersmith for 
particularly helpful suggestions 
regarding the condensed matter literature.
This work was supported by DOE grant DE-FG02-90ER-40560
and NSF grant PHY 91-23780.

\newpage
{\Large{{\bf Appendices}}}
\vskip 1cm

\appendix

\section{Calculation of the potential}

We need to evaluate
\bb \label{VV}
\mathcal{V}\equiv \lk< \frac{
\lk\{(\del_+X_r)^2 (\del_-X_r)^2-\lk[ (\del_+X_r)^2+(\del_-X_r)^2\re] 
(\del_+X_r.\del_-X_r)
\re\}}{\lk(X_r^2\re)^{D/2}}\re>\ ,
\ee
in the finite temperature vacuum of the SYM.
Let subscripts $(123456)$ denote the argument of $X$, \eg\
$X_1\equiv X(\sigma_1)$.
Writing $X_r\equiv X_5-X_6$, we will encounter in the numerator
only factors of the form
\bb \label{ugly1}
\del_\alpha X^i_1 \del_\alpha X^i_2 \del_\beta X^j_3 \del_\gamma X^j_4\ ,
\ee
with the target indices $i$, $j$ summed over; 
$\alpha$, $\beta$, $gamma$ are worldsheet indices $\pm$;
and the labels $(1234)$ are set equal to $5$ and $6$ in various ways.
By expanding the numerator of equation~\pref{VV}, 
we get $3\times 16$ terms of the form claimed.
We can write our desired `monomial'~\pref{ugly1} as
\bb \label{monom}
\del_\alpha^1 \del_\alpha^2 \del_\beta^3 \del_\gamma^4
\lk< \frac{X_1^{i} X_2^{i} X_3^{j} X_4^{j}}{((X_5-X_6)^2)^{D/2}} \re>\ .
\ee

Consider
\bbb 
\lk< \frac{X_1^{i} X_2^{i} X_3^{j} X_4^{j}}{((X_5-X_6)^2)^{D/2}} \re>
&=& \frac{\pi^{-d/2}}{\Gamma(D/2)} 
\int_0^\infty ds\ \int d^dp\ s^{(D/2)-1} e^{-p^2}
\delta_1^i\delta_2^i \delta_3^j 
\delta_4^j \lk <e^{\int \tilde{J}.X}\re > \nonumber\\
&=& \frac{\pi^{-d/2}}{\Gamma(D/2)} 
\int_0^\infty ds\ \int d^dp\ s^{(D/2)-1} e^{-p^2}
\delta_1^i \delta_2^i \delta_3^j \delta_4^j e^{\Delta}
\eee
where
\bb
\tilde{J^i}\equiv J^i+2 i \sqrt{s} 
(\delta(\sigma-\sigma_5)-\delta(\sigma-\sigma_6))p^i
\ee
and
\bb
\Delta\equiv \frac{1}{4} \int \tilde{J} K \tilde{J}=
\frac{1}{4} \int J K J+i\sqrt{s} \int J.p K_x-2sp^2 f^2
\ee
We have defined
\bb
K_x\equiv K_{x5}-K_{x6}
\ee
\bb
f^2\equiv K-K_{56}
\ee
Here $K_{ab}$ means $K(a-b)$, the Green's function of the two dimensional
Laplacian
\bb
K_{ab}\equiv -\alpha' \vev{X_a X_b}\ ,
\ee
and $K\equiv K_{aa}$. The rest is an exercise in combinatorics,
making use of
\bb
\delta_a^i e^\Delta= \lk[ \frac{1}{2} 
\int G_{ax} J^{i}+i\sqrt{s} p^i K_a\re] e^\Delta
\ee
where the $x$ subscript is integrated over 
and it is implied to be the argument of the $J$ as well.
Denoting the number of polarizations in the Lorentz indices by $d$, we get
\bbb
\lk< \frac{X_1^{i} X_2^{i} X_3^{j} X_4^{j}}{((X_5-X_6)^2)^{D/2}} \re>
&=&\frac{\pi^{-d/2}}{\Gamma(D/2)} \int ds \int d^dp\ e^{-p^2} e^{-2sp^2f^2} 
\nonumber \\
& &\hskip 1cm\times
\lk[ T_1 s^{(D/2)-1}-\frac{1}{2}p^2 s^{D/2} 
T_2+(p^2)^2 s^{(D/2)+1} T_4 \re]
\eee
where we have defined
\bb
T_1\equiv \frac{d^2}{4} K_{12} K_{34}+\frac{d}{4} 
K_{13}K_{24} +\frac{d}{4} K_{23} K_{14}
\ee
\bb
T_2\equiv d K_1 K_2 K_{34}+d K_3 K_4 K_{12}+
K_1K_3 K_{24}+K_1K_4K_{23}+K_2K_3K_{14}+K_2K_4K_{13}
\ee
\bb
T_4\equiv K_1 K_2 K_3 K_4
\ee
Evaluating the $s$ integral, we get
\bb
\lk< \frac{X_1^{i} X_2^{i} X_3^{j} X_4^{j}}{((X_5-X_6)^2)^{D/2}} \re>
=\frac{\pi^{-d/2}}{2^{D/2} (f^2)^{D/2}}\int d^dp\ e^{-p^2}(p^2)^{-D/2}
\lk[ T_1-\frac{D}{8f^2} T_2+\frac{(D+2)D}{16 (f^2)^2} T_4 \re]
\ee
Evaluating the $p$ integrals, we get
\bb
\lk< \frac{X_1^{i} X_2^{i} X_3^{j} X_4^{j}}{((X_5-X_6)^2)^{D/2}} \re>
=\frac{\pi^{-d/2}}{2^{(D/2)+1} (f^2)^{D/2}}\Omega_{d-1}
\Gamma\lk(\frac{d-D}{2}\re) \lk[
T_1-\frac{D}{8f^2} T_2+\frac{(D+2)D}{16 (f^2)^2} T_4 \re]
\ee
where $\Omega_{d-1}$ is the volume of the $d-1$ unit sphere.

Going back to~\pref{monom}, we need to differentiate $T_1$, $T_2$ and $T_4$,
according to the map $1234\rightarrow \alpha \alpha \beta \gamma$. Let us
denote the derivatives by superscripts on the $K$'s. We then have
\bb\label{T1a}
T_1^{\alpha \alpha \beta \gamma}=\frac{d^2}{4} K_{12}^{\alpha \alpha}
K_{34}^{\beta \gamma}+\frac{d}{4} 
K_{13}^{\alpha \beta} K_{24}^{\alpha \gamma}
+\frac{d}{4} K_{23}^{\alpha \beta} K_{14}^{\alpha \gamma}\ ,
\ee
\bb
T_2^{\alpha \alpha \beta \gamma}=
d K_1^\alpha K_2^\alpha K_{34}^{\beta \gamma}
+d K_3^\beta K_4^\gamma K_{12}^{\alpha \alpha}+
K_1^\alpha K_3^\beta K_{24}^{\alpha \gamma}+K_1^\alpha 
K_4^\gamma K_{23}^{\alpha \beta}
+K_2^\alpha K_3^\beta K_{14}^{\alpha \gamma}+
K_2^\alpha K_4^\gamma K_{13}^{\alpha \beta}\ ,
\ee
\bb\label{T4a}
T_4^{\alpha \alpha \beta \gamma}=
K_1^\alpha K_2^\alpha K_3^\beta K_4^\gamma\ .
\ee
We have used here the translational invariance and 
evenness of the Green's function to interpret the
derivatives as differentiations with respect to the argument $i-j$ of the 
Green's functions (and therefore note some flip of signs); furthermore,
we assume that $K$, $K^\alpha$ 
and $K^{\alpha \beta}$ are zero, \ie\ because of
subtraction of the zero temperature limits, or throwing away bubble 
diagrams. This, it turns out, is not necessary 
for the potential we calculate,
since all expressions would have come out 
as differences, say $K^{\alpha \beta}_{12}
-K^{\alpha \beta}$; it is just convenient for notational purposes to
throw them out from the start. We also note the identities
$K^{\alpha \beta}_{12}=K^{\alpha \beta}_{21}$ 
and $K^\alpha_5=-K_{56}^\alpha= K^\alpha_6$. 
For each term in equations~\pref{T1a}-\pref{T4a}, we have
16 terms associated with taking a map from 
$(1234)$ to a sequence of $5$'s and
$6$'s. This combinatorics yields
\bb
\lk< \frac{
\del_\alpha X_r^i \del_\alpha X_r^i \del_\beta X_r^j \del_\gamma X_r^j
}{\lk(X_r^2\re)^{D/2}}\re> \simeq
\frac{K_{56}^{\alpha \beta} K_{56}^{\alpha \gamma}+
\frac{d}{2} K_{56}^{\alpha \alpha} 
K_{56}^{\beta \gamma}}{(-K_{56})^{D/2}}\ .
\ee
Note that $T_2$ and $T_4$ cancelled; 
we have also dropped numerical coefficients.
There are three terms in~\pref{VV} of this type; this yields 
\bb
\mathcal{V}\simeq \frac{K_{56}^{+-} K_{56}^{+-}
+\frac{d}{2} K_{56}^{++} K_{56}^{--}
-\lk( \frac{d}{2}+1 \re) \lk( K_{56}^{++} K_{56}^{+-}
+K_{56}^{--} K_{56}^{+-} \re)
}{(-K_{56})^{D/2}}\ .
\ee
Using the equation of motion 
(delta singularity subtracted) $K_{56}^{+-}=0$,
we get
\bb
\mathcal{V}\simeq \frac{K_{56}^{++} K_{56}^{--}}{(-K_{56})^{D/2}}\ .
\ee
Using Euclidean time $i \tau=t$, 
we have $\sigma^\pm=\sigma\pm t=z,\bar{z}$;
finally, we get for equation~\pref{slinky}
\bb
V_{12} = \alpha_D g_s^2 \frac{\lstr^{5-p}}{R_+}
\frac{K^{zz}_{12} K^{\bar{z}\bar{z}}_{12}}{\lk( -K_{12} \re)^{D/2}} .
\ee

\section{Longitudinal momentum transfer effects}

Consider the scattering of two wound strings in IIB theory with winding
number exchange.  We will find that, in the regime of small 
momentum transfer, the interaction
is Coulombic for resonances involving
low enough winding number exchange,
and much weaker otherwise; furthermore, the Coulombic
interaction is winding number independent, and the cumulative
strength of this potential suggests modifying
the Matrix string potential by a factor of $N/S$ for $N>S$.

For simplicity, consider the polarizations of the external states
to be that of the dilaton, and T-dualize the momentum in the compact
direction to winding number.
The resulting four string amplitude is given by~\cite{GSW}
\bb
A_m\sim K_{\alpha \beta \gamma \delta} K^{\alpha \beta \gamma \delta}
\frac{\Gamma(-S\alpha'/4)\Gamma(-T\alpha'/4)\Gamma(-U\alpha'/4)}
{\Gamma(1+S\alpha'/4) \Gamma(1+T\alpha'/4)\Gamma(1+U\alpha'/4)} \ ,
\ee
where $S\equiv -(k_1+k_2)^2$, $T\equiv -(k_2+k_3)^2$, $U\equiv -(k_1+k_3)^2$,
with $S+T+U=0$, and 
\bbb
K^{\alpha \beta \gamma \delta}&=&-\frac{1}{2}\lk(
S T \eta^{\alpha \gamma} \eta^{\beta \delta} + S U \eta^{\beta \gamma}
\eta^{\alpha \delta}+ T U \eta^{\alpha \beta} \eta^{\gamma \delta} \re)
\nonumber \\
&+&S \lk( k_4^\alpha k_2^\gamma \eta^{\beta \delta}+
k_3^\beta k_1^\delta \eta^{\alpha \gamma} + k_3^\alpha k_2^\delta
\eta^{\beta \gamma} + k_4^\beta k_1^\gamma \eta^{\alpha \delta} \re)
\nonumber \\
&+&T \lk( k_4^\gamma k_2^\alpha \eta^{\beta \delta}+
k_3^\delta k_1^\beta \eta^{\alpha \gamma} + k_4^\beta k_3^\alpha
\eta^{\gamma \delta} + k_1^\gamma k_2^\delta \eta^{\alpha \beta} \re)
\nonumber \\
&+&U \lk( k_2^\alpha k_3^\delta \eta^{\beta \gamma}+
k_4^\gamma k_1^\beta \eta^{\alpha \delta} + k_4^\alpha k_3^\beta
\eta^{\gamma \delta} + k_2^\gamma k_1^\delta \eta^{\alpha \beta} \re)\ .
\eee
This gives the amplitude
\bb
A_m\sim \lk[ (S+T)^4+S^4+T^4\re] 
\frac{\Gamma(-S\alpha'/4)\Gamma(-T\alpha'/4)\Gamma(-U\alpha'/4)}
{\Gamma(1+S\alpha'/4) \Gamma(1+T\alpha'/4)\Gamma(1+U\alpha'/4)}\ .
\ee
We want to accord winding 
$n_1$, $n_2$, $n_3$ and $n_4$
to the four strings, on a circle of radius $R$;
without any momenta along this cycle, we can extract easily this process
from the amplitude above by
\bb\label{seq}
s=S+ M^2\ ,
\ee
\bb\label{teq} 
t=T+m^2\equiv-q^2\ ,
\ee
with 
\bb
M^2\equiv \lk( \frac{R (n_1+n_2)}{\alpha'}\re)^2\ ,
\ee
\bb
m^2\equiv \lk( \frac{R (n_3-n_2)}{\alpha'}\re)^2\ .
\ee
For large $m_1,m_2$, and small $m$, $q^2$ is the 
spatial momentum transfer between the strings in the
center-of-mass frame.  Thus $M\gg m$, and we are in
the non-relativistic regime $E_{cm}^2\gg q^2$.
From equations~\pref{seq} and \pref{teq}, 
we see that $S\gg T$.  Using this and
the identities $\Gamma(z) \Gamma(1-z) \sin
(\pi z )=\pi$ and $\Gamma(1+z)=z \Gamma(z)$, one obtains the amplitude
\bb\label{ampl}
A_m\sim (s-M^2)^2 
\sin \lk(\pi (q^2+m^2) \alpha'/4\re) \lk(\Gamma((q^2+m^2)
\alpha'/4)\re)^2\ .
\ee
In the energetic regime considered, 
\bb
s-M^2\sim m_1 m_2 v_{rel}^2\equiv \sqrt{\TT}\ ,
\ee
where $v_{rel}$ is the relative velocity of strings $1$ and $2$
in the lab frame.

Equation~\pref{ampl} has poles at $q^2+m^2=4 n/\alpha'$ with $n\le 0$.
We consider scattering processes probing distances $r$ much larger than
the string scale, $q_{max}\sim 1/r\ll 1/\lstr$; we also assume that 
it is possible to have
$R\ll \lstr$, which we will see is necessary. Given that these poles space the 
masses of the resonances by the string scale, the dominant term to
the amplitude is the one corresponding to the exchange of a wound ground
state, \ie\ the $n=0$ pole. 
Measuring quantities in string units, the amplitude then becomes
\bb
A_m\sim \TT \frac{\sin (q^2+m^2)}{\lk(
q^2+m^2\re)^2}\ .
\ee

The effective potential between the strings is the Fourier
transform of this expression with respect to $q$. 
Let us consider various limits.
Take $m \ll q$; we then have $m\ll 1$. The amplitude becomes
\bb
A_m^{(1)}\sim \frac{\TT}{q^2}\ .
\ee
Next consider $m\gg q$, but $m\ll 1$. The amplitude becomes
\bb
A_m^{(2)}\sim \frac{\TT}{q^2+m^2}\ .
\ee
Finally, for $m\gg q$ and $m\gg 1$, we have a constant
\bb
A_m^{(3)}\sim \TT \frac{\sin m^2}{m^4}\ .
\ee

The effective potentials are then ($d\equiv 9-p$)
\bb
V_{eff}^{(1)}\sim  \int d^d q\ e^{iq.x} A_m^{(1)} 
\sim  \frac{T}{r^{d-2}}\ .
\label{Vone}
\ee
The result is a Coulomb potential, independent of $m$.
The second case gives
\bb
V_{eff}^{(2)} 
\sim  T \lk(2\pi\re)^{d/2} \lk(\frac{m}{r}\re)^{d/2-1}
\sqrt{\frac{\pi}{2 m r}} e^{-mr} \ ,
\label{Vtwo}
\ee
which is weaker than $V_{eff}^{(1)}$ since we have $mr\gg 1$.
Finally, we have
\bb
V_{eff}^{(3)} 
  \sim T \frac{\sin m^2}{m^4}\frac{1}{r^d} \ .
\label{Vthree}
\ee
In addition to a larger power in $r$,
we have $m\gg 1$; this interaction is much weaker than
\pref{Vone},\pref{Vtwo}, especially after averaging
over a range of winding transfers $m$.

We conclude that, for $mr=R r(n_3-n_2)/\alpha' \ll 1$, 
we have a Coulombic potential independent
of the winding exchange $m$; for $mr\gg 1$, we have
much weaker potentials. 
This implies that in a gas of winding strings
bound in a ball of size at most of order the string scale,
the dominant potential is Coulombic with a multiplicative 
factor given by $w_0\equiv \alpha'/(R r)$,
provided a mechanism restricts winding exchange processes to 
$n_3-n_2\ll n_1+n_2$. 

The S-dual of this amplitude describes the scattering of wound
D-strings at strong coupling, with winding number exchange.
Under a further T duality, and lifting to M theory,
this amplitude encodes a good measure of the effects of longitudinal
momentum exchange in the problem of a self-interacting matrix string.
The bound on the winding number
translates in our language to
\bb
w_0=\frac{{\bar{\alpha}}' \bar{g}_s}{\Sigma r}=\frac{R_{11}}{r}\sim
\frac{N}{S}\ ,
\ee
\ie\ the resolution in the longitudinal direction. We also note that,
under this chain of dualities, the string scale used to set
a bound
on the impact parameter $r$ transforms as $\alpha'\rightarrow \alpha'$,
where the latter string scale is that of the matrix string. This justifies
our implied equivalence between the scale of
$r$ and that of the size of the black hole.

In the single matrix string case we study, 
we saw that regions of size $N/S$ were
strongly correlated and `rigid' in a statistical sense. 
The self-interaction of the large string will then involve processes of
coherent exchange of D-string
winding up to the winding number $N/S\ll N$. 
For larger winding, the D-string is not coherent; one
expects a suppression both from the emission vertex and 
from the highly off-shell propagator.
We saw above that all such processes, up to $N/S$, 
are of equal strength and scale Coulombically.
This implies that the potential between the string strands calculated
from the DBI expansion must be enhanced by a factor of $N/S$ for $N>S$, 
and justifies the scaling arguments used in Section 5.4.

\end{document}